\RequirePackage[colorlinks,allcolors=blue]{hyperref}
\documentclass[draft]{agujournal2019}
\usepackage{url}\usepackage{lineno}
\usepackage{soul}
\usepackage{amsmath}
\usepackage{siunitx}
\usepackage{nomencl}
\usepackage{cleveref}
\usepackage[super]{nth}
\usepackage{subcaption}
\usepackage[percent]{overpic}
\usepackage{blindtext}
\usepackage{amsfonts}
\usepackage{amssymb}
\usepackage{booktabs}
\usepackage{makecell}
\usepackage{lmodern}
\usepackage{graphicx}
\usepackage{ragged2e}
\justifying
\newcommand{\icon}{ICON}
\newcommand{\climsim}{ClimSim}
\newcommand{\climsimconv}{ClimSim Convection}
\newcommand{\radscheme}{``RTE+RRTMGP''}
\newcommand{\esmvaltool}{ESMValTool}
\newcommand{\kaggfirst}{``greySnow''}
\newcommand{\kaggsecond}{``Z Lab''}

\newcommand{\kaggfifth}{``YA HB MS EK''}

\draftfalse

\journalname{Journal of Advances in Modeling Earth Systems (JAMES)}

\begin{document}

\title{Beyond the Training Data: Confidence-Guided Mixing of Parameterizations in a Hybrid AI-Climate Model}

\authors{Helge Heuer\affil{1}, Tom Beucler\affil{2,3}, Mierk Schwabe\affil{1}, Julien Savre\affil{1}, Manuel Schlund\affil{1}, Veronika Eyring\affil{1,4}}

\affiliation{1}{Deutsches Zentrum für Luft- und Raumfahrt, Institut für Physik der Atmosphäre, Oberpfaffenhofen, Germany}
\affiliation{2}{Faculty of Geosciences and Environment, University of Lausanne, Lausanne, Switzerland}
\affiliation{3}{Expertise Center for Climate Extremes, University of Lausanne, Lausanne, Switzerland}
\affiliation{4}{University of Bremen, Institute of Environmental Physics (IUP), Bremen, Germany}

\correspondingauthor{Helge Heuer}{helge.heuer@dlr.de}

\begin{keypoints}
\item An ML convection parameterization trained on ClimSim and coupled to ICON achieves stable and accurate 20-year AMIP simulations.
\item Physics-informed loss, confidence-guided mixing, and noise-augmented training enhance conservation, accuracy, and stability, respectively.
\item The scheme can be tuned with observations by mixing in the conventional scheme when NN confidence is low in moist, unstable regimes.
\end{keypoints}

\begin{abstract}
Persistent systematic errors in Earth system models (ESMs) arise from difficulties in representing the full diversity of subgrid, multiscale atmospheric convection and turbulence. Machine learning (ML) parameterizations trained on short high-resolution simulations show strong potential to reduce these errors. However, stable long-term atmospheric simulations with hybrid (physics + ML) ESMs remain difficult, as neural networks (NNs) trained offline often destabilize online runs. Training convection parameterizations directly on coarse-grained data is challenging, notably because scales cannot be cleanly separated. This issue is mitigated using data from superparameterized simulations, which provide clearer scale separation. Yet, transferring a parameterization from one ESM to another remains difficult due to distribution shifts that induce large inference errors. Here, we present a proof-of-concept where a ClimSim-trained, physics-informed NN convection parameterization is successfully transferred to ICON-A. The scheme is (a) trained on adjusted ClimSim data with subtracted radiative tendencies, and (b) integrated into ICON-A. The NN parameterization predicts its own error, enabling mixing with a conventional convection scheme when confidence is low, thus making the hybrid AI-physics model tunable with respect to observations and reanalysis through mixing parameters. This improves process understanding by constraining convective tendencies across column water vapor, lower-tropospheric stability, and geographical conditions, yielding interpretable regime behavior. In AMIP-style setups, several hybrid configurations outperform the default convection scheme (e.g., improved precipitation statistics). With additive input noise during training, both hybrid and pure-ML schemes lead to stable simulations and remain physically consistent for at least 20 years, demonstrating inter-ESM transferability and advancing long-term integrability.

\end{abstract}

\section*{Plain Language Summary}
Clouds and thunderstorms are difficult to simulate accurately in climate models because they typically occur at scales smaller than the model’s grid. This necessitates the use of approximations for these processes, so-called parameterizations, which often introduce errors. Machine learning (ML) offers a new way to improve these models, but ML can be unstable and doesn’t always behave well when employed in different models or with different conditions.
In this study, we develop a new hybrid method that combines machine learning with established physical principles to better simulate the influence of atmospheric convection. Our approach learns from high-fidelity climate simulations and can adjust its behavior based on how confident the ML model is in its predictions.
This helps the model stay stable and accurate, even when it is used in a different climate model. Furthermore, a small amount of noise is added during training to improve the long-term stability of our ML model. We tested our method in the ICON climate model and found that it is accurate and stable in year-long simulations, while remaining stable and reliable over periods of 20 years. This work shows that blending physics with machine learning can lead to more accurate and robust climate models.

\section{Introduction}\label{sec:intro}
Mass-flux parameterization schemes, which represent the vertical transport of energy, water, and momentum in convective up- and downdrafts as a function of environmental conditions \cite{arakawa_interaction_1974,tiedtke_comprehensive_1989}, remain the de facto standard for parameterizing deep convection in modern ESMs. Such parameterizations can however introduce substantial biases into climate projections \cite{judt_insights_2018,stevens_dyamond_2019,christopoulos_assessing_2021,lee_future_2021} because they are often based on empirical relationships and simplifying assumptions.

Recent years have seen a surge of machine learning (ML)-based parameterizations for deep convection and cloud physics \cite{gentine_could_2018,yuval_stable_2020,yuval_neural-network_2023,heuer_interpretable_2024}. Training ML-based schemes on coarse-grained high-resolution data and implementing them in conventional Earth System Models (ESMs) promises to reduce long-standing biases in coarse-scale global simulations. To design a suitable training dataset, the choice of the coarse-graining and filtering operator is however critical and not uniquely defined \cite{ross2023benchmarking,brenowitz_interpreting_2020}. Furthermore, coarse-graining storm-resolving \icon{} data does not yield a clean separation of convective versus other subgrid processes \cite{heuer_interpretable_2024}. Additionally, generating global storm-resolving training data is extremely expensive \cite{satoh_global_2019} and most available storm-resolving ICON datasets are not ideal as a ground truth because they do not offer an appropriate temporal output frequency (sub-hourly), global coverage, or do not include the needed variables for training: DYAMOND \cite{stevens_dyamond_2019} or nextGEMS \cite{koldunov2023nextgems} provide only 3-hourly 3D fields and at best 15-min 2D surface variables; NARVAL \cite{stevens2019AHighAltitudeLongRangeAircraftConfiguredasaCloudObservatoryTheNARVALExpeditions,klocke2017rediscovery} is confined to the tropical Atlantic with hourly output. Challenges related to using complex high-resolution training data are illustrated in our previous work, \citeA{heuer_interpretable_2024}, where an ML model for deep convection was trained on coarse-grained and filtered two-month-long high-resolution tropical data. This yielded promising online results such as an improved representation of precipitation extremes, but also introduced heavy blurring and biases in variables such as column water vapor or temperature.

Furthermore, whereas previously developed ML parameterizations have shown success in modeling the subgrid convective fluxes and convective precipitation, stability issues remain very common, even in idealized aquaplanet setups \cite{gentine_could_2018,rasp_deep_2018,brenowitz_prognostic_2018,brenowitz_spatially_2019,brenowitz_interpreting_2020,yuval_stable_2020,lin2025navigating}. Hybrid ML–physics climate models have yet to demonstrate stable, accurate simulations suitable for operational use; emerging real-geography runs are still too short \cite{watt-meyer_neural_2024} or too coarse \cite{hu2025stable,wang2022StableClimate,han2023EnsembleNeural}.

In an attempt to mitigate the discussed challenges of training ML models on global storm-resolving data directly, we use the \climsim{} dataset \cite{JMLR:v26:24-1014}, generated with the Energy Exascale ESM multiscale modeling framework (E3SM-MMF) \cite{e3sm_project_energy_2018}. In this superparameterized setup \cite{hannah_initial_2020}, 2D Storm Resolving Models (SRMs) with periodic boundaries are embedded in each coarse atmospheric column, replacing conventional subgrid parameterizations. \climsim{} pairs coarse-scale atmospheric states (inputs) with tendencies derived from the embedded SRMs (targets), providing a well-defined, although artificial \cite{hannah_checkerboard_2022}, scale separation between resolved coarse dynamics and unresolved physics. This reduces ad hoc choices in coarse-graining and process separation when training ML models. In addition, a 2024 challenge on Kaggle, an open ML competition platform, built around \climsim{}, attracted over 690 final submissions \cite{lin_leap_2024}, yielding strong baselines and architectures we leverage here.

In this proof-of-concept, we leverage these developments to create a new ML-based parameterization of convection for the Icosahedral Nonhydrostatic (ICON) model \cite{giorgetta_icon-_2018,zangl_icon_2015} with a horizontal resolution of $\qty{\sim 160}{\kilo\meter}\times\qty{\sim 160}{\kilo\meter}$, trained on the \climsim{} dataset. Our ML approach draws inspiration from models developed in the Kaggle competition, in which vertically recurrent neural networks (NNs) \cite{ukkonen2025verticallyrecurrentNN}, such as bi-directional Long Short-Term Memory (BiLSTM) architectures, emerged as competitive contenders for predicting subgrid-scale tendencies from large-scale inputs. We additionally implement a physically informed loss function encouraging the trained networks to adhere to conservation laws and to discourage non-conservative sources and sinks in single-column predictions. A key modification, inspired by the first-place winner \kaggfirst{} of the Kaggle competition, is the incorporation of a confidence loss. This adds a second prediction head that estimates the loss for all targets, effectively quantifying model uncertainty. Using this confidence metric during online inference, we mix ML predictions with the conventional convection scheme when the ML scheme is uncertain, thereby improving overall performance. The approach is similar to the novelty-detection method of \citeA{sanford_improving_2023} or the ``compound parameterization'' proposed by \citeA{krasnopolsky_neural_2008} and used in \citeA{song2021compound}, identifying and responding to out-of-distribution or uncertain conditions during inference. Rather than applying ML corrections unconditionally, we use the confidence metric as a proxy for uncertainty to detect potential extrapolation beyond the training domain. By avoiding extrapolation and applying ML corrections only in specific regions of the input space, this method prevents unphysical or biased outputs and enhances stability and reliability. While the method by \citeA{sanford_improving_2023} might be more reliable in out-of-distribution cases, the use of the confidence metric is easier to implement and use as it does not require a separate novelty detection algorithm. With this work, we build upon previous studies demonstrating ML-based parameterizations in ICON \cite{grundner_deep_2022,grundner_data-driven_2024,grundner_reduced_2025,heuer_interpretable_2024,hafner_interpretable_2024,sarauer2025physics}.

This paper is organized as follows: \Cref{sec:Data} presents the \climsim{} dataset used for model training, along with the observational and reanalysis datasets for evaluation. \Cref{sec:ml_scheme_method} outlines the overall methodology, detailing the architecture of the ML-based convection scheme, the loss design, and the confidence-guided mixing. In \Cref{sec:results}, we evaluate one-year-long coupled simulations, analyzing climate statistics and the physical behavior of the ML parameterization to gain process-level insights. As a comprehensive validation, we conduct historical Atmospheric Model Intercomparison Project (AMIP)-type simulations with prescribed sea surface temperatures (SSTs), sea-ice concentrations, and greenhouse gas concentrations. Finally, \Cref{sec:conclusion} discusses the key findings and concludes the study.

\section{Data}\label{sec:Data}

\subsection{ClimSim and Cross-Validation Procedure}
We used the ``high-resolution version'' of the \climsim{} dataset \cite{yu_climsim_2023,ai_climsim_high-res_2023} with a horizontal resolution of approximately $1.5^{\circ}\times1.5^{\circ}$. The data are produced over realistic geography with E3SM-MMF \cite{e3sm_project_energy_2018}, span 2005–2014 with \qty{20}{\min} output, and total about \qty{41.2}{\tera\byte} \cite{JMLR:v26:24-1014}. Sea surface temperatures and sea-ice amount were prescribed. Boundary conditions such as ozone and aerosol concentrations were set to the climatological average of 2005–2014 \cite{JMLR:v26:24-1014}. In this multiscale modeling framework, subgrid-scale dynamics are resolved by 2D SRMs embedded within each grid column of the coarse atmospheric model. These SRMs have a horizontal resolution of \qty{2}{\kilo\meter} and are two-way coupled to the coarse atmospheric model \cite{hannah_checkerboard_2022}. The SRMs replace the coarse model’s parameterizations for convection and boundary-layer turbulence \cite{lee_representation_2023} and are used for the calculation of radiative fluxes. The SRMs are mostly based on the System for Atmospheric Modeling (SAM; \citeA{Khairoutdinov2003CloudResolvingModelingoftheARMSummer1997IOPModelFormulationResultsUncertaintiesandSensitivities}), use SAM’s single-moment microphysics, and close sub-SRM-grid-scale turbulent fluxes with a diagnostic Smagorinsky-type closure. Gravity wave drag and vertical diffusion are parameterized by the coarse atmospheric model outside the SRMs \cite{JMLR:v26:24-1014}. We refer the curious reader to \citeA{yu_climsim_2023} for more details on \climsim{} and \citeA{hannah_checkerboard_2022} for the E3SM-MMF setup.

This dataset offers several advantages compared to training data from other high-resolution models that enhance its utility for research.
Notably, it features a well-defined scale separation between subgrid-scale and grid-scale dynamics, as it is generated through a superparameterized modeling framework.
Additionally, the dataset is readily accessible to the research community and was utilized in a Kaggle competition \cite{lin_leap_2024} that attracted over 690 finalized submissions.
The collaborative efforts of participants in this competition have yielded highly competitive machine learning models and baselines, providing a valuable benchmark for future studies and inspiring innovative approaches to data-driven modeling.

Potential drawbacks of learning from the superparameterized \climsim{} data set are the usage of 2D SRMs with limited extent for the embedded subgrid dynamics and the useful but artificial scale separation \cite{hannah_checkerboard_2022}. Therefore, the subgrid dynamics are highly idealized and can, e.g., influence the mean state response affecting moisture and associated shortwave cloud effects \cite{pritchard2014restricting}. Additionally, as shown later in \Cref{sec:results:benchmarking_with_obs} and \Cref{sec:results:20_year_amip}, the zonal precipitation distribution of the high-res version of \climsim{} shows too high mean precipitation with respect to the Global Precipitation Climatology Project (GPCP), especially in the mid to high latitudes as well as for the Intertropical Convergence Zone (ITCZ).

To train ML models efficiently while utilizing the temporal variability of the data we only used the first two days of every month over the span of the ten years with a timestep of \qty{20}{\min}. This resulted in approximately \num{217e6}/\allowbreak{}\num{37e6}/\allowbreak{}\num{37e6} training/\allowbreak{}validation/\allowbreak{}test samples.
For more efficient training of the NNs we further subsampled the data in time and space, ending up with a \num{25e6}/\allowbreak{}\num{5e6}/\allowbreak{}\num{5e6} training/\allowbreak{}validation/\allowbreak{}test split.

\subsection{``\climsimconv{}'': Approximate Removal of Radiation for Training}\label{sec:data:radremoval}

\begin{figure}[tbh]
    \centering
       \includegraphics[width=1\linewidth]{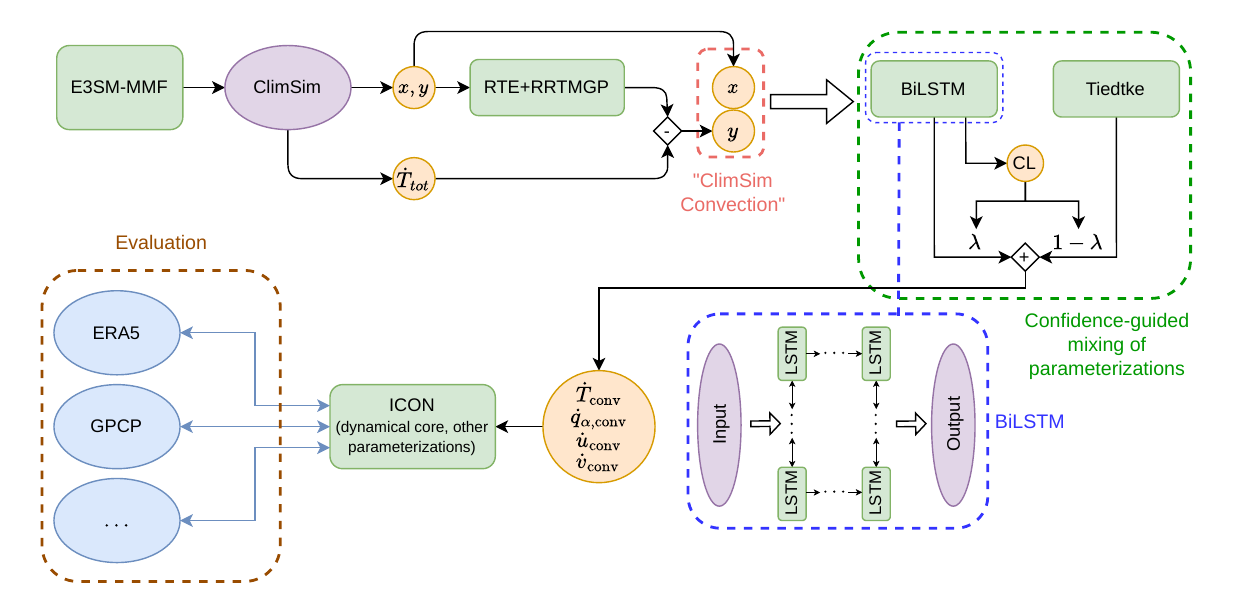}
    \caption{Overall training and evaluation pipeline of our hybrid model. $x$ and $y$ represent inputs and outputs of the \climsim{} dataset, based on the E3SM-MMF model. $\dot{T}_{tot}$ is the total temperature tendency, and \radscheme{} the ICON radiation scheme. The \climsim{} dataset is first modified to separate radiative and convective subgrid tendencies, forming a new dataset, ``\climsimconv{}''. Afterward, we trained a BiLSTM model including a confidence loss (CL). Using CL, this model is mixed with the conventional ``Tiedtke'' cumulus convection scheme to predict convective tendencies as well as precipitation. In the mixing process, $\lambda$ represents the fraction provided by the BiLSTM and ${1-\lambda}$ is the fraction from the conventional ``Tiedtke'' scheme, respectively. This mixed scheme predicts the tendencies due to convection in temperature $\dot{T}_\mathrm{conv}$, water vapor, cloud liquid water, cloud ice ($\dot{q}_{\alpha,conv}$, $\alpha={v,l,i}$), zonal wind $\dot{u}_\mathrm{conv}$, and meridional wind $\dot{v}_\mathrm{conv}$. Finally, we coupled the mixed scheme with the ICON model and evaluate these runs' emergent statistics with respect to observational datasets, including ERA5 and GPCP.}
    \label{fig:method_vis}
\end{figure}

While \climsim{} facilitates process separation, it does not cleanly isolate convective processes. To use ClimSim to train a drop-in replacement for ICON's convection scheme, we must avoid double-counting radiation. Addressing this is an essential step toward creating a modular version of \icon{} where different model components can be run using either their conventional or data-driven schemes. We therefore constructed \climsimconv{} by subtracting radiative temperature tendencies from \climsim{}. Because the E3SM-MMF subgrid state is unavailable, we approximated the radiative contribution by recomputing column radiation offline with the \radscheme{} scheme \cite{pincus_balancing_2019,pincus_rterrtmgp_2023}. This scheme was employed to generate the \climsim{} dataset \cite{yu_climsim-online_2024,lee_representation_2023} and is also used in our ICON setup. The resulting radiative heating is subtracted from the superparameterized temperature tendencies (see \Cref{fig:method_vis}).

\radscheme{} is driven by per-column inputs from \climsim{}: temperature; tracers for specific humidity, cloud liquid, and cloud ice; solar insolation and the solar zenith angle's cosine; ozone, N$_2$O, and CH$_4$; shortwave/longwave albedos; surface pressure; and outgoing longwave radiation. Mid- and half-level pressures are reconstructed from the time-independent coefficients \texttt{hyam}/\texttt{hybm} and \texttt{hyai}/\texttt{hybi} provided by \climsim{}:

\begin{align}
\begin{split}
    P_{m,k} = \texttt{hyam}_k P_0 + \texttt{hybm}_k P_\mathrm{sfc},\qquad
    P_{h,k} = \texttt{hyai}_k P_0 + \texttt{hybi}_k P_\mathrm{sfc},
\end{split}
\end{align}

with $P_0=\qty{1000}{\hecto\pascal}$, and $k$ representing the height level index. Cloud effective radii are computed as in ICON.

This subtraction yields tendencies dominated by convective heating, which we aim to learn, with residual contributions from microphysics and turbulence. Explicitly separating convection from microphysics/turbulence would be ad hoc and arguably unphysical \cite{arakawa_cumulus_2004,arakawa_multiscale_2011,randall_breaking_2003}. Accordingly, in coupled runs we replaced only deep convection in ICON and keep its native vertical diffusion and microphysics scheme active; once radiation was removed, we found no evidence of residual double-counting (e.g., anomalous diffusion signatures; not shown). Furthermore, we set up \icon{} simulations without vertical diffusion and/or without microphysics schemes. These simulations became numerically unstable almost immediately. We note, however, that because \climsim{} also includes non‑convective precipitation \cite{liu2023UnderstandingPrecipitation}, there is a possibility of double counting non‑convective rainfall within \icon{}.

\Cref{fig:rterrtmgp_dTdt_distributions} in \Cref{sec:app:additional_figures} shows that removing radiative tendencies preserves the distributional shape across the column and yields a net convective heating (left) that balances the removed longwave cooling (middle), with shortwave heating as expected (right), consistent with the atmospheric energy budget. Overall, \climsimconv{} keeps assumptions minimal while acknowledging \climsim{}’s imperfections when training convective parameterizations. Learning from \climsimconv{} is therefore treated as a transfer-learning exercise between E3SM-MMF and \icon{}, which requires online validation.%In what follows, we compare the hybrid scheme against ICON’s conventional Tiedtke convection using ERA5 and GPCP.

\subsection{Datasets Used for Evaluation}\label{sec:data_eval}

For the evaluation of the coupled \icon{} online runs, we mainly employed two datasets: GPCP \cite{adler_global_2018} and the ERA5 reanalysis \cite{hersbach_era5_2020}. The GPCP dataset provides a comprehensive, long-term record of global precipitation, combining various satellite observations, rain gauge measurements, and other remote sensing data. GPCP offers a spatial resolution of \qty{2.5}{\degree} × \qty{2.5}{\degree} and temporal coverage spanning several decades with monthly temporal resolution, making it ideal for validating simulated precipitation patterns against observational benchmarks. ERA5, on the other hand, is the fifth-generation ECMWF reanalysis dataset, which provides atmospheric data at a higher resolution than GPCP (about \qty{0.25}{\degree} × \qty{0.25}{\degree}) at hourly intervals. It incorporates a wide range of variables, including temperature, wind, humidity, and surface pressure, and is widely used for evaluating climate models due to its high accuracy and consistency with physical laws.
These datasets were chosen for their broad applicability, high quality, and availability, enabling a direct and meaningful evaluation of the model's performance in real-world scenarios.

For the bulk of the evaluation, we used the Earth System Model Evaluation Tool (\esmvaltool{}) \cite{righi_earth_2020,andela_esmvaltool_2025}. \esmvaltool{} is a community diagnostic and performance metrics tool for the evaluation of Earth system models (ESMs) \cite{righi_earth_2020}.
Besides the ERA5 and GPCP references, \esmvaltool{} offers the possibility to evaluate against a multi-observational mean for certain variables. These datasets additionally include, e.g., MERRA2 \cite{gelaro_modern-era_2017}, ESACCI-WATERVAPOUR \cite{schroder_combined_2023}, and ISCCP-FH \cite{zhang_global_2023}. We used the multi-observational mean to evaluate the spatial distribution of column integrated water vapor.
For evaluating precipitation statistics, we utilized the GPCP dataset, while ERA5 is used for the near-surface temperature $T_{2m}$.

\section{Parameterization Schemes Methodology}\label{sec:ml_scheme_method}

This section describes how the conventional Tiedtke scheme compares to our newly developed ML-based scheme. After introducing the Tiedtke scheme, we outline the methodology behind training the ML model, including constructing its loss function, selecting hyperparameters, and implementing confidence-guided mixing in ICON.

As seen in \Cref{fig:method_vis}, we used the \climsimconv{} data to train NNs (with a physics informed loss) to predict the convective tendencies and convective precipitation with the atmospheric state variables as input. The NNs are based on a bidirectional long short term memory (BiLSTM) architecture and trained with a confidence loss inspired by the first place entry \kaggfirst{} to the \climsim{} Kaggle competition \cite{lin_leap_2024}. This enables the networks to judge their own prediction error during inference.
We leveraged these error predictions for a mixed convection parameterization in which the NNs' predictions are mixed with those from the conventional cumulus convection scheme when the NNs exhibit low confidence, as explained in \Cref{sec:method:confmix}.%We called the fraction of the mixed model stemming from the ML predictions $\lambda$, which means $\left(1-\lambda\right)$ is the fraction used for the conventional cumulus convection scheme. The mixing was performed in a continuous manner as explained later in \Cref{sec:method:loss_func:confloss}.

\subsection{Tiedtke Convection Scheme}\label{sec:method:tiedtke}
As described in \citeA{giorgetta_icon-_2018,mobis_factors_2012}, the conventional cumulus convection scheme used in the \icon{} model is based on a mass flux formulation by \citeA{tiedtke_comprehensive_1989} with modifications by \citeA{nordeng_extended_1994}. It differentiates between shallow, mid-level, and deep convection. Deep convection occurs in disturbed environments with synoptic scale convergence whereas undisturbed environments allow for shallow convection \cite{tiedtke_comprehensive_1989}. Mid-level convection originates at levels above the boundary layer and is often formed by lifting of low level air until saturation \cite{tiedtke_comprehensive_1989,blanchard_mid-level_2021}. For deep convection, an adjustment-type closure based on the Convective Available Potential Energy is used. Shallow convection uses a moisture convergence closure and a large scale vertical momentum closure which determines the cloud base mass-flux for mid-level convection. The scheme represents all subgrid convective cloud processes by one updraft and one downdraft, respectively.

The bulk convection scheme works by defining a vertical profile for the mass-flux $M(z)$ which varies by the amount of entrainment and detrainment happening in the up-/downdrafts (for downdrafts only turbulent entrainment/detrainment is considered \cite{nordeng_extended_1994}). To determine the magnitude of the mass-flux and relate the subgrid convection process to the resolved large-scale flow, the three different closures are used. Tendencies for temperature, water vapor, cloud liquid water, cloud ice, and zonal/meridional wind are calculated with this scheme. The convective rain and snow rates are also computed and analyzed.
We refer to this scheme as ``Tiedtke scheme'' in this study.

\subsection{Machine Learning Scheme}\label{sec:method:mlscheme}
\begin{figure}[tbh]
    \centering
    \includegraphics[width=1\linewidth]{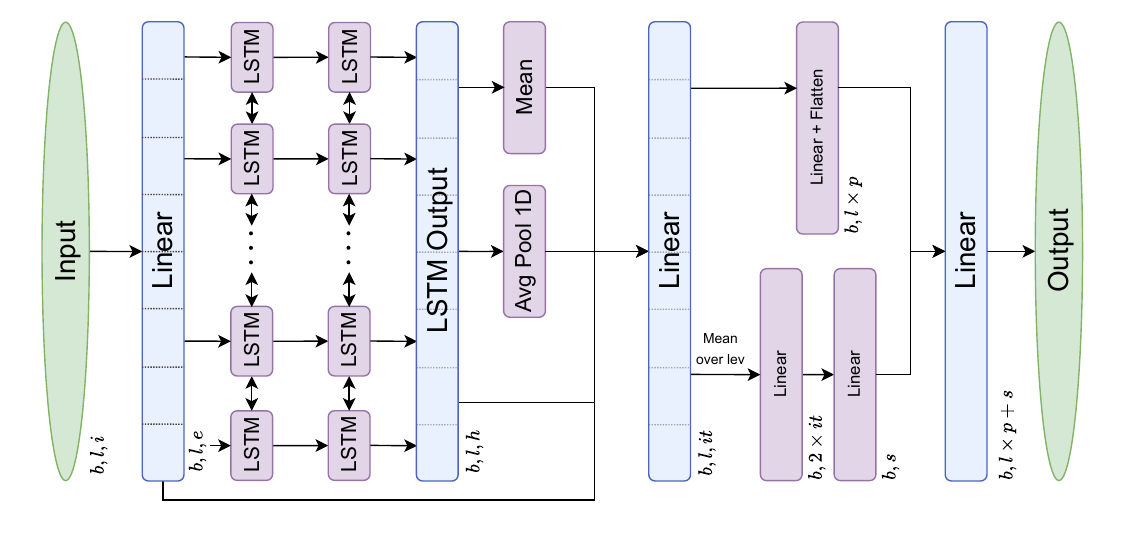}
    \caption{The BiLSTM architecture developed by the \nth{5} place Kaggle competition winner \kaggfifth{}, and used in the work presented in this article. Tensor dimensions are visualized in the lower right corner of the individual layers. The tensor dimensions shown in the figure are the batch dimension $b$, the column height level dimension $l$, the input dimension $i$, the encoding dimension $e$, hidden dimension $h$, iter dimension $it$, output scalar dimension $s$, and the output profile dimension $p$. In the blue-marked layers, the horizontal dotted lines indicate operations restricted to the last dimension, thereby preserving ``vertical locality''.}
    \label{fig:BilstmVis}
\end{figure}

The backbone architecture for the selected NN is a BiLSTM. Our implementation is a BiLSTM based on the winner of the \nth{5} place in the Kaggle competition, \kaggfifth{} \cite{lin_leap_2024}, and considers sequences along the model height dimension for each column. We selected this approach due to its accessibility and the demonstrated effectiveness of BiLSTMs in capturing vertical profiles for atmospheric parameterization tasks \cite{yao_physics-incorporated_2023,ukkonen_representing_2024,hafner_interpretable_2024}. Furthermore, in the Kaggle competition, the solution of the \nth{5} placed team only had a difference of 0.0037 in its coefficient of determination $R^2$ compared to the \nth{1} place (\kaggfirst{}) on the private leaderboard, which we do not expect to make a significant difference for the coupled online skill.
Our architecture is shown in \Cref{fig:BilstmVis}.

The numerical values of the various dimensions shown in \Cref{fig:BilstmVis} are given in \Cref{tab:HPOs}. The inputs to the ML model used in this work were inspired by \citeA{hu2025stable} and consist of the input variable set:

\begin{align*}
    I = \{&T,RH,q_\mathrm{l},q_\mathrm{i},\chi_\mathrm{liq},u,v,\dot{T}_{t-1},\dot{q}_{\mathrm{v},t-1},\dot{q}_{\mathrm{l},t-1},\dot{q}_{\mathrm{i},t-1},\dot{u}_{t-1},\\ &\dot{T}_{t-2},\dot{q}_{\mathrm{v},t-2},\dot{q}_{\mathrm{l},t-2},\dot{q}_{\mathrm{i},t-2},\dot{u}_{t-2}\},
\end{align*}

with temperature $T$, relative humidity $RH$, cloud liquid water $q_\mathrm{l}$, cloud ice $q_\mathrm{i}$, liquid partition $\chi_\mathrm{liq}$, zonal wind $u$, meridional wind $v$, and water vapor $q_\mathrm{v}$. All variables with a dot superscript are convective tendencies from the last (${t-1}$ subscript) or second to last (${t-2}$ subscript) timestep. The liquid partition $\chi_\mathrm{liq}$ is a function of the temperature and has a value of 1 for temperatures above \qty{0}{\degree\C} and 0 for temperatures below \qty{-20}{\degree\C}. Between \qty{-20}{\degree\C} and \qty{0}{\degree\C} the function varies linearly as shown in Figure 2 of \citeA{hu2025stable}.

This input set is similar to the inputs the conventional Tiedtke scheme uses, but also includes atmospheric variables from the two previous timesteps. The choice to include inputs from the timesteps $t-1$ and $t-2$ was also inspired by \citeA{hu2025stable} and can be motivated by the fact that by suppressing access to the high-resolution state, the evolution of the low-resolution state is conditionally dependent on the low-resolution states of previous timesteps as argued in \citeA{beucler_distilling_2025}. Furthermore, by incorporating information from previous time steps, especially from the thermodynamic variables temperature and water vapor, the scheme gains the capability to capture convective memory effects \cite{colin_identifying_2019}.

The model outputs the following set of variables:

\begin{equation*}
    O=\{\dot{T},\dot{q_\mathrm{v}},\dot{q_\mathrm{l}},\dot{q_\mathrm{i}},\dot{u},\dot{v},{\cal P}_{\mathrm{rain}},{\cal P}_{\mathrm{snow}}\},
\end{equation*}

with the two 2D variables convective rain rate ${\cal P}_{\mathrm{rain}}$ and convective snow rate ${\cal P}_{\mathrm{snow}}$. The other variables are 3D tendencies for temperature, water vapor, cloud liquid water, cloud ice, and zonal/meridional wind.

We implemented our NNs in PyTorch \cite{ansel_pytorch_2024} and PyTorch Lightning \cite{falcon_pytorch_2019}. Inspired by the Kaggle competition, we chose AdamW as the optimizer \cite{loshchilov_decoupled_2019}.

\subsection{Loss Function}\label{sec:method:loss_func}

\subsubsection{Total Loss}\label{sec:method:loss_func:totloss}

The total per-sample loss during training $\ell_{\mathrm{tot}}$ combines the Huber loss $\ell_\mathrm{Huber}$ with the confidence loss $\ell_\mathrm{conf}$, the ``difference'' loss $\ell_\mathrm{diff}$, and a physics-informed loss $\ell_{\varphi}$ grouping the residual of the enthalpy, mass, and momentum budgets. These terms are explained in the following subsections, and the overall loss is computed as

\begin{equation}
\ell_{\mathrm{tot}}(\hat{y},y) = \alpha \, s \cdot \ell_{\varphi}\left(x,\hat{y}\right) + (1-\alpha) \cdot \left[\ell_{\mathrm{Huber}}(\hat{y},y)  + \ell_\mathrm{diff}(\hat{y},y)+ \ell_\mathrm{conf}\left( \hat{y}_\mathrm{loss},\hat{y},y\right)\right].
\label{eq:loss_function}
\end{equation}

The parameter $\alpha$ serves as a tunable hyperparameter that governs the relative weight of the physically informed loss terms.
To ensure an approximately equal contribution from both the data-driven and the physics-based components, we introduced another hyperparameter $s$. We initially trained the model without minimizing the physical residuals, instead quantifying their magnitude during this phase. Empirical analysis revealed that a scaling factor of approximately $s=385$ effectively balances the magnitudes of these terms. This factor was subsequently applied to the summed physical residuals prior to their integration into the overall loss function, thereby enabling stable and effective backpropagation during subsequent training iterations.

\subsubsection{Huber Loss}\label{sec:method:loss_func:huber}

As the Huber loss and other combinations of the $L_1$ and $L_2$ loss terms were used successfully by many teams in the Kaggle competition, we chose the Huber loss with hyperparameter $\delta=1$ as our base loss. An $L_2$ loss is applied for absolute biases between predictions $\hat{y}$ and targets $y$ smaller than $\delta$, and an $L_1$ loss otherwise:

\begin{equation}
        \ell_{\mathrm{Huber}}(\hat{y},y) = \begin{cases}
        0.5\cdot (y - \hat{y})^2, & \text{if } |y - \hat{y}| < \delta \\
        \delta\cdot \left(|y - \hat{y}| - 0.5 \cdot \delta\right), & \mathrm{otherwise}.
        \end{cases}
\end{equation}

\subsubsection{Physics-Informed Loss}\label{sec:method:loss_func:physicsloss}

The physical loss $\ell_{\varphi}$ is introduced to reduce enthalpy, mass, and momentum conservation errors in the ML scheme during training. Note that the conventional scheme in \icon{} strictly conserves these quantities in the vertical or converts atmospheric water phases to precipitation, whereas the \climsim{} dataset does not, because processes like radiation and land surface fluxes lead to net inflows and outflows of conserved quantities.
For numerical stability and ease of implementation, we implemented the calculation of the physical terms in the BiLSTM architecture in non-dimensional form. The constants we chose for non-dimensionalization are the following:

\begin{align*}
    g_0&=\qty{9.80665}{\metre\per\s\tothe{2}},\\
    t_0&=\qty{10}{\s},\\
    \rho_\mathrm{h2o}&=\qty{1000}{\kg\per\metre\tothe{3}},\\
    c_{p}&=\qty{1004.64}{\joule\per\kelvin\per\kg}.
\end{align*}

The choice of these scales was physically motivated and their numerical values were taken from the \icon{} model, except for the timescale, which was chosen so that the derived scales for, e.g., length, energy, temperature, and pressure are reasonably close to statistical average values of the dataset. Non-dimensional variables are henceforth denoted with tildes and more details about the non-dimensionalization of the physical terms can be found in \Cref{sec:app:non-dim}.

Our physics-informed loss

\begin{equation}
\ell_{\varphi}=\widetilde{H}_\mathrm{res}+\widetilde{m}_\mathrm{res}+\widetilde{u}_\mathrm{res}+\widetilde{v}_\mathrm{res}
\end{equation}

sums the non-dimensional residual fluxes of conserved variables, which were calculated as follows:

\begin{align}
    \label{eq:e_residual}
    \widetilde{H}_\mathrm{res} &= \int_{\tilde{p}_\mathrm{top}}^{\tilde{p}_\mathrm{bot}} \left(\frac{\partial \widetilde{T}}{\partial \widetilde{t}} - \frac{\partial \widetilde{q_\mathrm{l}}}{\partial \widetilde{t}}\cdot \widetilde{L}_\mathrm{v} - \frac{\partial \widetilde{q_\mathrm{i}}}{\partial \widetilde{t}}\cdot \widetilde{L}_\mathrm{s}\right) d\tilde{p} - \widetilde{L}_\mathrm{v}\cdot \widetilde{\cal P}_{\mathrm{rain}} - \widetilde{L}_\mathrm{s}\cdot \widetilde{\cal P}_{\mathrm{snow}}, \\
    \label{eq:m_residual}
    \widetilde{m}_\mathrm{res} &= \int_{\tilde{p}_\mathrm{top}}^{\tilde{p}_\mathrm{bot}}  \left(\frac{\partial \widetilde{q_\mathrm{v}}}{\partial \widetilde{t}} + \frac{\partial \widetilde{q_\mathrm{l}}}{\partial \widetilde{t}} + \frac{\partial \widetilde{q_\mathrm{i}}}{\partial \widetilde{t}}\right) d\tilde{p} + \widetilde{\cal P}_{\mathrm{rain}} + \widetilde{\cal P}_{\mathrm{snow}}, \\
    \label{eq:u_residual}
    \widetilde{u}_\mathrm{res} &= \int_{\tilde{p}_\mathrm{top}}^{\tilde{p}_\mathrm{bot}}  \frac{\partial \widetilde{u}}{\partial \widetilde{t}}\ d\tilde{p}, \\
    \label{eq:v_residual}
    \widetilde{v}_\mathrm{res} &= \int_{\tilde{p}_\mathrm{top}}^{\tilde{p}_\mathrm{bot}}  \frac{\partial \widetilde{v}}{\partial \widetilde{t}}\ d\tilde{p}.
\end{align}

$\widetilde{L}_v$ and $\widetilde{L}_s$ are the non-dimensionalized latent heat of vaporization and sublimation.
The residual fluxes for the conserved quantities (enthalpy $\widetilde{H}_\mathrm{res}$, mass $\widetilde{m}_\mathrm{res}$, zonal momentum $\widetilde{u}_\mathrm{res}$, and meridional momentum $\widetilde{v}_\mathrm{res}$), were calculated following \Cref{eq:e_residual,eq:m_residual,eq:u_residual,eq:v_residual} by integration over the pressure coordinate, necessitating the inclusion of mid-level and surface pressure as inputs to the neural network.
In the integrals, the pressure coordinate ranges from the pressure level of the highest predicted level $\tilde{p}_\mathrm{top}$ to the surface pressure $\tilde{p}_\mathrm{bot}$. These pressure variables were utilized solely for computing differences between pressure half-levels within the model code, which were then employed in the residual flux calculation and were not used in the forward pass of the network itself.
\Cref{eq:e_residual,eq:m_residual} contain terms for $q_\mathrm{v}$, $q_\mathrm{l}$, and $q_\mathrm{i}$ only, as rain and snow are not treated as 3D resolved tracers in the setup of \icon{} and the convective parameterization respectively.

Adding these residual fluxes to the loss function in \Cref{eq:loss_function} effectively encouraged the model to redistribute the conserved quantities in a column instead of introducing non-physical sources or sinks. As a result, the NNs trained in this manner are no longer purely data-driven, but rather physics-informed.  

\subsubsection{Improving the Output's Vertical Structure via the ``Difference Loss''}\label{sec:method:loss_func:diffloss}

Inspired by the \nth{2} place (\kaggsecond{}) solution of the Kaggle competition \cite{lin_leap_2024}, to help the model learn the vertical structure of each predicted profile, we included an additional loss term $\ell_\mathrm{diff}(\hat{y},y)$ that quantifies the error between real and predicted differences of vertically adjacent levels:

\begin{equation}
    \ell_\mathrm{diff}(\hat{y},y) = \sum_{i=1}^{N_{\mathrm{lev}}-1} \ell_{\mathrm{Huber}}(\hat{y}_{i+1}-\hat{y}_i,y_{i+1}-y_i),
\end{equation}

where $i$ indexes the vertical level and $N_{\mathrm{lev}}$ is the total number of vertical levels.

\subsubsection{Confidence Loss}\label{sec:method:loss_func:confloss}

Finally, inspired by the first place solution of the Kaggle competition from \kaggfirst{} and the AlphaFold loss function \cite{jumper2021highly}, we implemented a technique in which the NN estimates its own prediction error. The method introduces a second prediction head by doubling the number of output neurons in the final layer, where the second half of the output layer predicts the error of the predictions $\hat{y}_\mathrm{loss}$.
Combining these loss predictions and minimizing the resulting ``confidence-loss'' term defined as:

\begin{equation}
    \ell_{\mathrm{conf}}\left( \hat{y}_\mathrm{loss},\hat{y},y\right) = \ell_{\mathrm{Huber}}\left(\hat{y}_\mathrm{loss},\ell_{\mathrm{Huber}}\left(\hat{y},y\right)\right) 
\end{equation}    

ensures that the network learns to estimate its own loss as accurately as possible. In practice, the model is able to anticipate when its predictive skill is reduced because of high variability in the output due to, e.g., latent drivers, or when predictions are made in regions of the input feature space containing few training samples.%Assuming monotonous behavior of the confidence estimate in directions of decreasing data density, the model also becomes able to judge when it is extrapolating.

\subsection{Confidence-Guided Mixing}\label{sec:method:confmix}

\begin{figure}[tbh]
    \centering
    \includegraphics{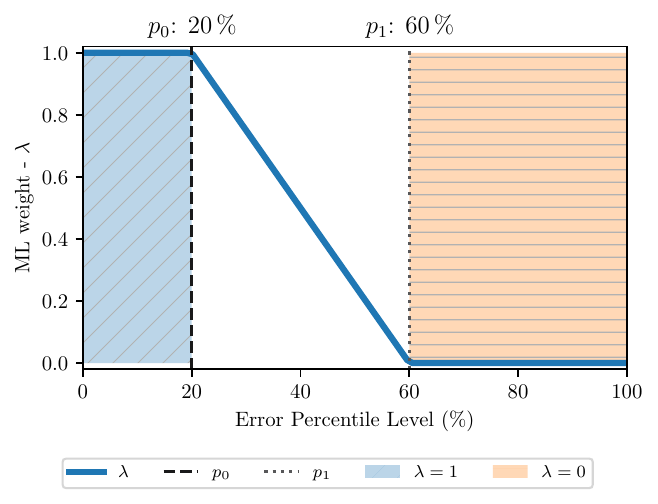}
    \caption{ML weight $\lambda$ as function of the predicted error percentile level. The tuning parameters $p_0$ and $p_1$ (here \qty{20}{\percent} and \qty{60}{\percent}) are marked by dashed and dotted lines, respectively. In blue and with slanted hatching, the area with $\lambda=1$ (pure ML) is shown. $\lambda=0$ (pure Tiedtke) is shown in orange and with horizontal hatching.}
    \label{fig:p020_p160_vis}
\end{figure}

On the validation set, we estimate the empirical cumulative distribution function (CDF) $F_{\mathrm{val}}$ of the predicted-loss averaged over all outputs. In practice, we store 101 equally spaced percentiles (\qtyrange{0}{100}{\percent}), which are used to approximate $F_{\mathrm{val}}$. In coupled runs, each online predicted error $\hat{y}_{\mathrm{loss}}$ is mapped to its percentile rank

\begin{equation}
    q = 100\,F_{\mathrm{val}}(\hat{y}_{\mathrm{loss}})\in[0,100].
\end{equation}

To ensure a smooth transition between the pure ML and conventional schemes, confidence-guided mixing uses two user-set percentile levels $p_0<p_1$ (e.g., 20 and 60), defined with respect to $F_{\mathrm{val}}$ (Fig.~\ref{fig:p020_p160_vis}). Expressing thresholds in percent makes them scale-free and comparable across models. Given $q$, the ML weight $\lambda$ is then defined as a linear ramp:

\begin{equation}
    \lambda(q)=\max\!\left\{0,\;\min\!\left[1,\;1-\frac{q-p_0}{p_1-p_0}\right]\right\}.
\end{equation}

Predicted tendencies are then mixed component-wise as

\begin{equation}
    \hat{y}_{\mathrm{mixed}}=\lambda\,\hat{y}+(1-\lambda)\,\hat{y}_{\mathrm{Tiedtke}}.
\end{equation}

Importantly, $F_{\mathrm{val}}$ (the mapping from error to percentile rank) is fixed from the validation set, while $p_0$ and $p_1$ offer the possibility to tune the coupled hybrid ICON model in order to better match observations; this avoids conflating the empirical percentiles with the mixing thresholds.

This confidence-guided mixing is coupled online to \icon{}, and the resulting tendencies are integrated with the model’s other parameterized and dynamical tendencies in the dynamical core \cite{zangl_icon_2015}.

Under a changing climate scenario, the ML weight $\lambda$ may decrease, leading to a greater contribution from the conventional parameterization in future climate simulations. This shift could introduce a change in the error characteristics of the simulated climate response, as the two schemes have different biases. However, this behavior is also a key strength of the method: by smoothly transitioning toward the conventional scheme in regions of phase space that are underrepresented or outside the ML model's training distribution, such as extreme or novel climate states, it avoids out-of-distribution failures that could otherwise degrade model performance. Given the infeasibility of generating training data for every possible future climate state, this adaptive robustness is a valuable feature.

\subsection{Jointly Optimizing Performance and Inference Cost}\label{sec:method:loss_func:skillcomplexitytradeoff}

\begin{figure}[tbh]
    \centering
    \includegraphics[width=1\linewidth]{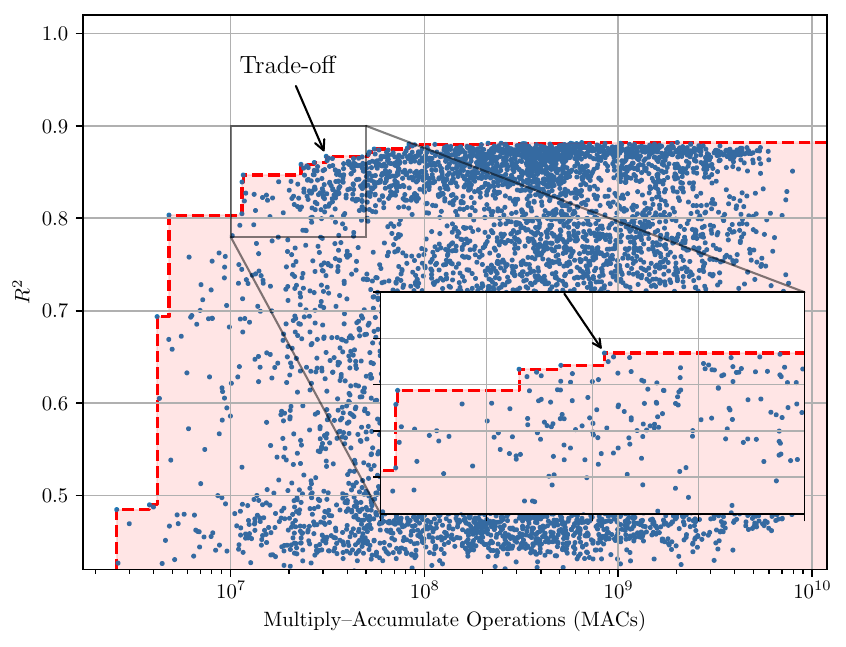}
    \caption{Offline skill-complexity plane for various combinations of nine chosen hyperparameters of the BiLSTM on a smaller subset of the dataset with \qty{3}{million} training and \qty{1.5}{million} validation samples. The red dashed line shows the Pareto Front between the coefficient of determination $R^2$ and the number of Multiply-Accumulate Operations (MACs). The highlighted NN is selected for the remainder of this study because it strikes a suitable balance between skill and computational performance.}
    \label{fig:offline_HPO_pareto}
\end{figure}

The original BiLSTM used by the \nth{5} place winner \kaggfifth{} in the Kaggle competition has around \qty{18}{million} trainable parameters.
To find a balance between model skill and computational efficiency, we first used \texttt{Ray Tune} \cite{liaw_tune_2018} on a smaller data subset of \qty{3}{million} training and \qty{1.5}{million} validation samples.
We varied the encoding dimension, the hidden dimension, the iteration dimension, the number of LSTM layers, and the dropout rate within the NN architecture.
For the optimizer/ scheduler we additionally varied the learning rate, the weight decay parameter, the batch size, and the type of scheduler.
The model marked as ''Trade-off`` in \Cref{fig:offline_HPO_pareto} has about \qty{540}{k} trainable parameters. This hyperparameter setting is used in the remainder of this study.
More information on the search space and the optimal parameters is given in \Cref{sec:app:opt}.

\Cref{fig:offline_HPO_pareto} shows all tested configurations and their coefficient of determination, as well as the number of Multiply-Accumulate Operations (MACs).
We also measured the inference time on the CPU directly for each of the models shown and found a correlation of $\sim \qty{99}{\percent}$ between MACs and inference time on CPUs, thus demonstrating that MACs are an appropriate measure of computational performance.The correlation of MACs with the GPU inference time is only $\sim\qty{9}{\percent}$, meaning that if we were doing such a skill-complexity comparison for a coupled model running on a GPU, we should look at the GPU inference time directly.
We decided to perform this comparison on the CPU instead of the GPU, as the NN is later coupled to the \icon{} model on the CPU.
This might change in the future as ICON can be run on GPUs \cite{giorgetta_icon-model_2022}.

The usefulness of Pareto fronts for ML models in climate modeling has been demonstrated in \citeA{beucler_distilling_2025}. Given multiple metrics, Pareto fronts are defined as the set of NNs for which no other NN $\cal M$ exists such that $\cal M$ shows an improvement in one metric without a worsening in any other metric relative to the original NN. Testing a limited number of other NN architectures along the Pareto front revealed that our results did not seem to be very sensitive to the specific architecture chosen (not shown).

\subsection{Additive Noise During Training for Improved Stability}\label{sec:methods:noise}

Inspired by the ``Engression'' framework by \citeA{shen_engression_2024} and by the results of \citeA{brenowitz_interpreting_2020}, we made the ML schemes more robust with respect to the transfer to a new domain with slightly different distributions. This method is applied only for the results presented in \Cref{sec:results:20_year_amip}. We did this by including additive noise to the inputs during training of the BiLSTMs:

\begin{equation}
    y = \mathrm{NN}\left(x+\eta\right),\quad\eta\sim\mathcal{N}\left(0,\sigma^2\right),
\end{equation}

where $\eta$ is a noise vector sampled from a Gaussian distribution $\mathcal{N}$ with zero mean and a tunable variance $\sigma^2$.
As $x$ and $y$ are standardized using a Z-score, the variance is constant across variables in $x$.
This preadditive noise can reveal some information about the true function outside the domain it was trained on, which can enable data-driven extrapolation \cite{shen_engression_2024}.

To add noise during training, we performed a warm restart from an optimized, noise-free NN. Algorithmically, we implemented a Python class initialized with four hyperparameters: the initial noise level $\sigma_0>0$; the tolerated $R^2$ drop compared to its value before any noise addition, $\Delta R^2>0$; and multiplicative growth/decay factors $m_\uparrow>1$ and $m_\downarrow\in(0,1)$ for the noise. In the first epoch, we add Gaussian input noise with standard deviation $\sigma_0$. After each epoch, we compute the change in $R^2$: if the drop exceeds $\Delta R^2$, we reduce the noise by $m_\downarrow$; otherwise, we increase it by $m_\uparrow$. After a manual search, we adopted $(\sigma_0,\Delta R^2,m_\uparrow,m_\downarrow)=\allowbreak{}(0.05,0.2,1.1,0.9)$.

\subsection{Online Coupling to ICON}\label{sec:methods_icon}

We used the \icon{} 2.6.4 model version with a horizontal resolution of approximately $\qty{158}{\kilo\meter}\times\qty{158}{\kilo\meter}$, corresponding to an \texttt{R2B4} \icon{} grid.
The model incorporates a range of parameterized processes, including radiation, cloud microphysics, orographic and non-orographic gravity wave drag, boundary layer turbulence, and convection.
Since our approach consists in mixing the pure ML and physical convection parameterizations, our ML-based model did not replace the original Tiedtke scheme but was run alongside it.
In order to initialize the convective tendencies of the two most recent timesteps needed by the ML convection scheme, we utilized the two last timesteps from the Tiedtke scheme as initial conditions.

To ensure compatibility between our ICON setup and the \climsim{} data, we configured \icon{} with 60 vertical levels, adjusting their heights to approximately match those of the \climsim{} dataset.
The ML schemes' tendencies were then coupled within the troposphere, and only the lowest 42 levels (corresponding to an approximate upper pressure level of \qty{95}{\hecto\pascal}) were used as inputs/outputs for the scheme.

For the coupling of the \icon{} model implemented in FORTRAN and the ML models in Python/\allowbreak{}PyTorch, we used the FTorch library \cite{atkinson_ftorch_2025}.
This library enables running the ML models in inference mode during the time integration of the \icon{} model.
After training and before exporting the ML models, we added normalization layers before and after the \texttt{forward} method of the model to take care of the preprocessing and postprocessing of the inputs and outputs during inference.
Coupling an ML model to ICON slows down the overall simulation by a factor of about two.

\section{Results}\label{sec:results}

This section first compares \icon{} simulations coupled to the various ML schemes developed in this study and the conventional Tiedtke scheme with observations. These comparisons use \esmvaltool{} \cite{righi_earth_2020,andela_esmvaltool_2025} to calculate evaluation metrics. We then examine the conservation properties of the developed models and investigate under which conditions they exhibit higher or lower confidence. Additionally, we explore why the mixed model demonstrates better skill than both the Tiedtke and pure ML models to ensure the improvements to convective physics are interpretable. Finally, this section concludes with an application of the developed schemes in 20-year-long AMIP-style simulations.

\subsection{Benchmarking with Observations}\label{sec:results:benchmarking_with_obs}

To evaluate the online performance of various ML models, we systematically varied the weight of the physics-informed loss term, $\alpha$, during training, with $\alpha\in\{0,\allowbreak{}0.01,\allowbreak{}0.1,\allowbreak{}0.5,\allowbreak{}0.9\}$.
The offline coefficients of determination on the test set for the models with $\alpha\leq 0.5$ are approximately $R^2\approx 0.89$ and $R^2=0.631$ for $\alpha=0.9$ as seen in \Cref{tab:offlineR2}.
Furthermore, we explored the impact of adjustments to the percentile parameters $p_0$ and $p_1$, which generated diverse ML weight configurations, $\lambda$.
Specifically, we tested $p_1$ values within the range of \qty{20}{\percent} to \qty{90}{\percent}, while $p_0$ was varied between \qty{10}{\percent} and $p_1$.
Additionally, we evaluated a model without the proposed mixing mechanism and no physics-informed loss terms ($\alpha=0$), referred to as the ``pure ML'' model, to establish a further baseline for comparison.
The simulations were run in an AMIP-style setup over an entire year starting on January \nth{1} 2010.
First, we will evaluate the performance of the ML-based schemes on some key climate metrics mainly related to water vapor and precipitation as the representation of water in the atmosphere is crucial for improving current climate models \cite{stevens_what_2013}.

\begin{figure}[tbh]
  \centering
  \begin{subfigure}[t]{0.49\textwidth}
    \vskip 0pt
    \begin{overpic}[width=\textwidth]{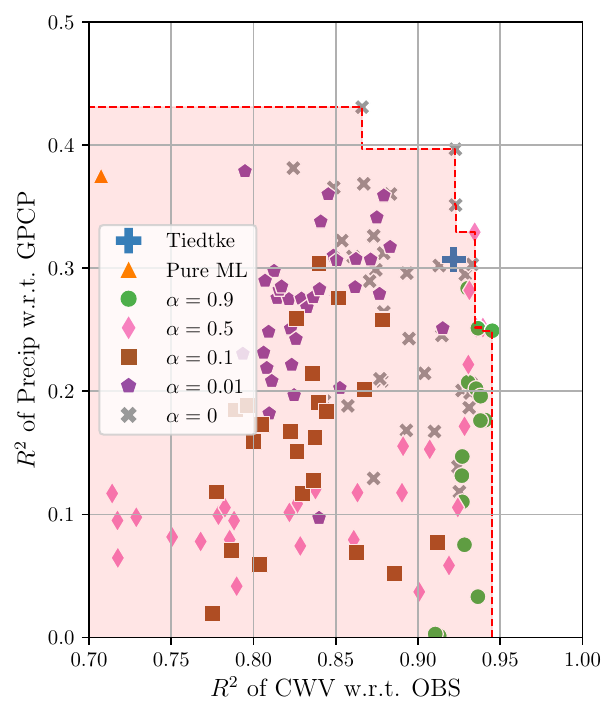}
      \put(0,100){a)}
    \end{overpic}
  \end{subfigure}
  \hfill
  \begin{subfigure}[t]{0.49\textwidth}
    \vskip 0pt
    \begin{overpic}[width=\textwidth]{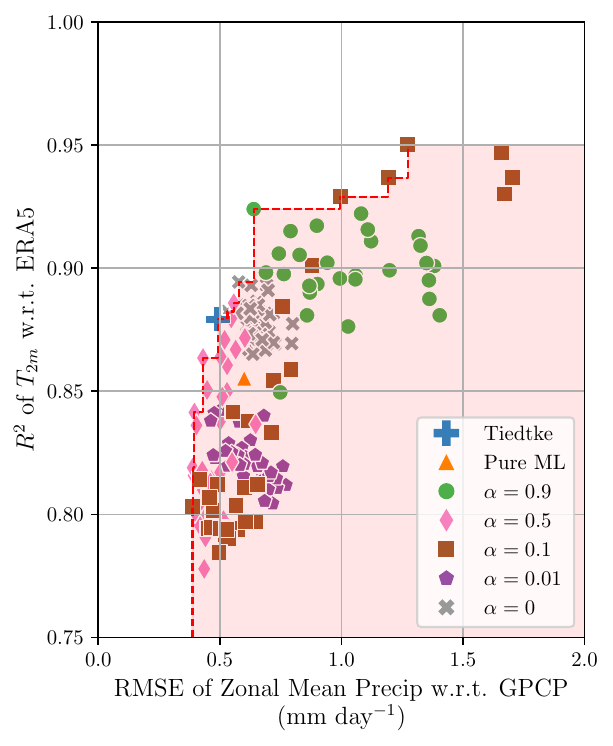}
      \put(0,100){b)}
    \end{overpic}
  \end{subfigure}
  \caption{Evaluation scores for coupled ICON runs, each dot represents a one-year long coupled ICON run at a horizontal resolution of $\qty{158}{\kilo\meter}\times\qty{158}{\kilo\meter}$. The runs are colored according to their physics-informed loss weight $\alpha$ for the coupled ML schemes and the conventional Tiedtke scheme is colored in blue. Within each coloring group, the models have different values for $p_0$ and $p_1$. Panel (a) shows the spatial $R^2$ score of precipitation with respect to the observational dataset GPCP versus the $R^2$ score of column water vapor (CWV) with respect to the mean of multiple observation sets as explained in \Cref{sec:data_eval}. Panel (b) displays the $R^2$ score of near-surface (\qty{2}{\meter}) air temperature with respect to ERA5 versus the RMSE of zonal mean precipitation with respect to GPCP. In both panels, the Pareto front between the two skill metrics is marked with a dashed red line. All metrics shown are based on the temporal averages of the variables over the one-year-long simulation period and are calculated over all longitudes and latitudes for the respective variable.}
  \label{fig:online_pareto_plots}
\end{figure}

\Cref{fig:online_pareto_plots} shows the performance of various model configurations evaluated by four different online metrics using \esmvaltool{}.
The conventional Tiedtke scheme is located near the Pareto front in panel (a) and on the Pareto front for (b).
This is not surprising, as the ICON model has been tuned to perform well when used with the default Tiedtke convection scheme.
Nevertheless, many coupled ML schemes lie along the Pareto front and we could expect even better results if \icon{} was calibrated with these schemes, which is not feasible for all of them.
In panel (a), we find a model with an $\alpha$ parameter of 0.5, showing an increase of $\Delta R^2\approx 0.015$ relative to the Tiedtke scheme in both metrics. Interestingly, some schemes outperform the Tiedtke scheme by a large margin with respect to one metric but have a lower skill in another metric.
For example, there is a model with $\alpha=0$ having a precipitation $R^2$ increase of over $\sim 0.12$ compared to Tiedtke and a scheme with $\alpha=0.9$ showing a column water vapor (CWV) $R^2$ increase of $\sim 0.25$. On panel (b), a clearer ordering of the $\alpha$ parameter with respect to the two metrics is observed.
Furthermore, panel (b) demonstrates that there exist ML schemes outperforming the Tiedtke scheme by $\sim 0.075$ in near-surface (\qty{2}{\meter}) air temperature $R^2$ and $\sim\qty{0.12}{\milli\metre\per\day}$ RMSE of the zonal mean precipitation.

The mixed models are named in the format ``Mixed:$p_0$-$p_1$\_\,$x\alpha$'', with $x$ indicating the value of the physics-informed loss weight $\alpha$.
The mixed model with the parameter setting $\{p_0=10,p_1=60,\alpha=0.1\}$ shows the least error of zonal mean precipitation with respect to observations and is therefore used for further analysis. For some of the shown results in this section, this model is compared to models having $\alpha=0.1$ as well, but different confidence parameters $p_0,p_1$.
For ease of notation, we will therefore leave out the $\alpha$ parameter in the naming of the model whenever $\alpha=0.1$.

\begin{figure}[tbh]
  \centering
  \begin{subfigure}[b]{0.49\textwidth}
    \begin{overpic}[width=\textwidth]{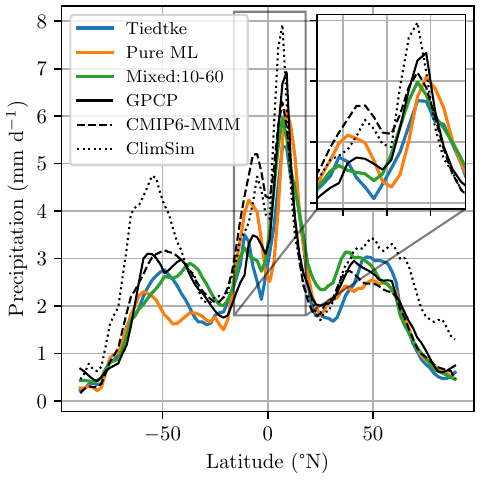}
      \put(0,100){a)}
    \end{overpic}
  \end{subfigure}
  \hfill
  \begin{subfigure}[b]{0.49\textwidth}
    \begin{overpic}[width=\textwidth]{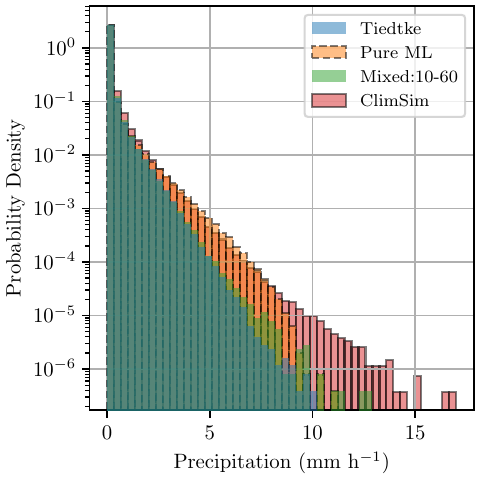}
      \put(0,100){b)}
    \end{overpic}
  \end{subfigure}
  \caption{Zonal mean precipitation in one-year-long runs (a) and precipitation distribution (b) for the pure ML scheme, the Tiedtke scheme, a mixed scheme (Mixed:10-60), and references; GPCP observations, CMIP6 multi-model mean (MMM), and \climsim{} (E3SM-MMF) for the zonal mean precipitation (a) as well as for the precipitation extremes (b).}
  \label{fig:precipitation_zonal_and_extremes}
\end{figure}

We next analyze the representation of precipitation in the various models by looking at zonal means of annual surface precipitation (\Cref{fig:precipitation_zonal_and_extremes}).
The Tiedtke scheme significantly underestimates the peak in mean precipitation (\Cref{fig:precipitation_zonal_and_extremes} (a)).
The pure ML scheme exhibits a stronger peak, although it remains lower than the GPCP reference.
The mixed scheme yields values slightly below the pure ML scheme, yet it outperforms the Tiedtke model.
The displayed Mixed:10-60 scheme represents a model ``tuned'' to observations as it shows the least RMSE of the tested model with respect to zonal mean precipitation of GPCP.
Notably, both the Tiedtke and pure ML schemes clearly display a signature of a double ITCZ in the sense that they show a pronounced second precipitation peak in the Southern Hemisphere.
The double ITCZ is however substantially less pronounced in the mixed scheme and more closely resembles the observational reference.
In the high latitudes all schemes exhibit a similar behavior.

To investigate the double ITCZ bias more quantitatively, we use the tropical precipitation asymmetry index $A_P$ \cite{hwang_link_2013} and the equatorial precipitation index $E_P$ \cite{adam_relation_2016}.
The tropical precipitation asymmetry index quantifies the asymmetry of tropical precipitation, with positive values indicating higher precipitation in the northern ($\qty{0}{\degree}-\qty{20}{\degree N}$) tropical hemisphere $\bar{P}_{\qty{0}{}-\qty{20}{N}}$ vs. the southern ($\qty{20}{\degree S}-\qty{0}{\degree}$) tropical hemisphere $\bar{P}_{\qty{20}{S}-\qty{0}{}}$ (and vice versa for negative values):

\begin{equation}
    A_P = \frac{\bar{P}_{\qty{0}{}-\qty{20}{N}} - \bar{P}_{\qty{20}{S}-\qty{0}{}}}{\bar{P}_{\qty{20}{S}-\qty{20}{N}}}.
\end{equation}

The equatorial precipitation index represents the symmetric component of tropical precipitation by relating the mean precipitation within \qty{2}{\degree S} - \qty{2}{\degree N}, $\bar{P}_{\qty{2}{S}-\qty{2}{N}}$, to the mean precipitation estimated between the tropics, $\bar{P}_{\qty{20}{S}-\qty{20}{N}}$:

\begin{equation}
    E_P = \frac{\bar{P}_{\qty{2}{S}-\qty{2}{N}}}{\bar{P}_{\qty{20}{S}-\qty{20}{N}}}.
\end{equation}

The respective biases are defined as the index for a model run minus the index evaluated for the observations.

\begin{table}[htb]
    \centering
    \begin{tabular}{l l l l l l}
        \toprule
        Data & $A_P$ & $E_P$ & $A_P$ Bias & $E_P$ Bias & RMSE (\unit{{\milli\meter\per\day}})\\
        \midrule
        GPCP & 0.454 & 0.920 & - & - & - \\
        \hline
               Tiedtke & 0.417 & 0.848 & -0.037 & -0.072 & 0.491\\
        Pure ML & 0.253 & 0.716 & -0.201 & -0.204 & 0.600\\
        Mixed:10-60 & 0.451 & 0.911 & \textbf{-0.003} & \textbf{-0.009} & \textbf{0.387}\\
        \hline
        \climsim{} & 0.268 & 0.973 & -0.186 & 0.053 & 0.884 \\
        CMIP6-MMM & 0.060 & 1.037 & -0.394 & 0.117 & 0.525\\
        \bottomrule
    \end{tabular}
    \caption{The tropical precipitation asymmetry index $A_P$ and the equatorial precipitation index $E_P$, and their biases, as well as the RMSE, with respect to GPCP for the data shown in \Cref{fig:precipitation_zonal_and_extremes} (a).}
    \label{tab:itcz_indices_one_year}
\end{table}

As \Cref{tab:itcz_indices_one_year} shows, the double ITCZ bias is lowest for the mixed model while the Tiedtke and pure ML models have significantly higher biases.
The informative value of these indices is rather limited due to their simplicity, but they give a further indication that the mixed model captures the zonal precipitation distribution well.
The mixed model also displays the lowest error as indicated by the RMSE of the curve of zonal mean precipitation with respect to the GPCP curve (\Cref{tab:itcz_indices_one_year}).

The distributions of daily precipitation values (Panel (b) of \Cref{fig:precipitation_zonal_and_extremes}) reveal notable differences between the various datasets.
The \climsim{} dataset stands out with the highest extreme precipitation values, which is expected given that it is based on the MMF data.
In contrast, the Tiedtke scheme underestimates precipitation extremes compared to \climsim{} and exhibits an overabundance of minor precipitation events, a phenomenon commonly known as the ``drizzle problem'' \cite{stephens_dreary_2010,wang_stochastic_2016}.
The ML scheme presents a distribution more akin to \climsim{} but appears to slightly overemphasize mid-level precipitation events, specifically those ranging from \qty{2}{\milli\metre\per\hour} to \qty{9}{\milli\metre\per\hour}.
Meanwhile, the mixed scheme offers a balance between low and high precipitation events, showcasing slightly more heavy precipitation events than the Tiedtke scheme, although still falling short of replicating the reference data provided by \climsim{}.

As a comparison to Figure 14 of \citeA{heuer_interpretable_2024}, we also visualize three snapshots of the column water vapor for some of the tested configurations. This is shown in \Cref{sec:app:additional_figures} (\Cref{fig:cwv_monthly_evolution}). In \citeA{heuer_interpretable_2024} a significant smoothing for the stable simulations was visible after 4 days and after one month there were no structures visible in the troposphere anymore. \Cref{fig:cwv_monthly_evolution} clearly shows that this is improved substantially as clear structures are still visible for all configurations after a month and even a year of integration.

\subsection{Advantages of Physics-Informed Loss via Conservation Laws}\label{sec:results:physlossconservation}

\begin{figure}[tbh]
    \centering
    \includegraphics[width=1\linewidth]{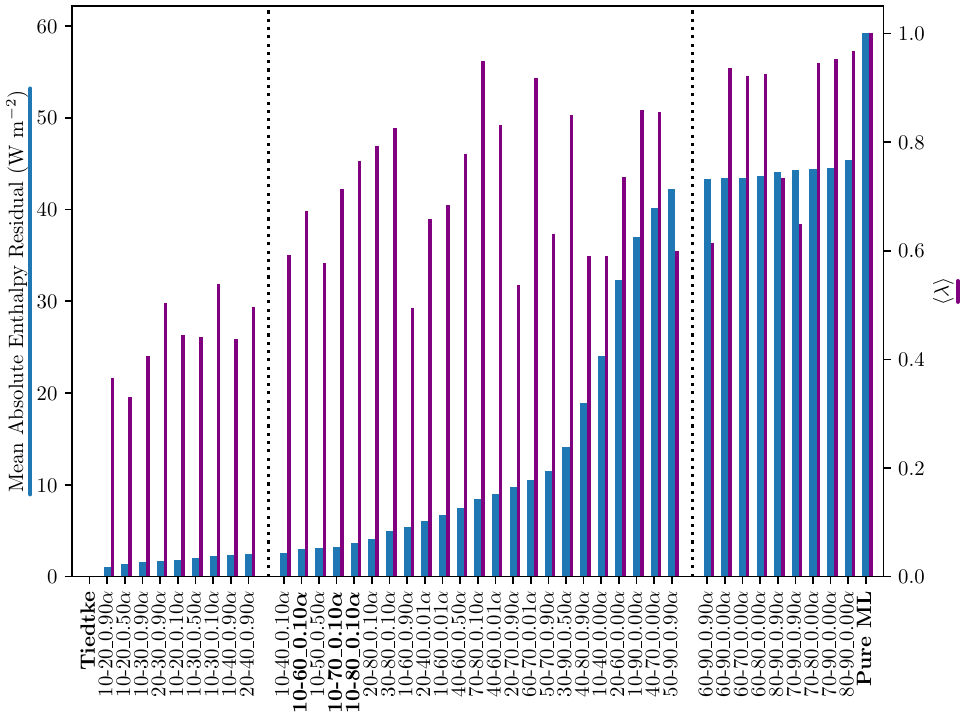}
    \caption{Mean absolute enthalpy residual (blue, left axis) and average ML weight $\lambda$ during the one-year long online integration (purple, right axis) for a selection of tested models. The ten most-conserving (left in the plot) and least-conserving (right) models in terms of enthalpy conservation are displayed. In between the black dotted lines every \nth{8} model is displayed so that the figure is still readable. Additionally, models which are used for a deeper analysis in this section are marked by bold labels.}
    \label{fig:enthalpy_residuals_online}
\end{figure}

To assess the fidelity of the learned physics, we monitor the mean absolute enthalpy residual, i.e., the mean absolute value of \Cref{eq:e_residual}, throughout the simulations, alongside the global mean ML weight, $\langle\lambda\rangle$ (\Cref{fig:enthalpy_residuals_online}).
As expected, the conventional Tiedtke scheme demonstrates perfect enthalpy conservation.
Conversely, the pure ML scheme exhibits the largest residuals as it has learned no physical conservation laws during training and also does not mix in any conservative Tiedtke output profiles.
Notably, the NNs enforcing soft constraints on enthalpy, mass, and momentum conservation, exhibit intermediate behavior.
This demonstrates that the proposed hybrid approach effectively constrains the ML predictions, resulting in improved physical consistency compared to a purely data-driven model, which is particularly relevant for long-term integrations.

Assuming the loss distribution would stay the same when coupled to \icon{}, one would expect an average ML weight of $\langle\lambda\rangle = (p_0 + (p_1 - p_0)/2)/100$. However, in practice this is not the case as seen in \Cref{fig:enthalpy_residuals_online} and indicates that there is a considerable distributional shift between the training distribution and the one encountered during inference coupled to the ICON model.

\subsection{Process Understanding: Why is the Mixed Model Better Than Both the Tiedtke and Pure ML Model?}\label{sec:results:process_understanding}

In this section, we analyze the mixed scheme across environmental regimes defined by geography (latitude), CWV, and lower-tropospheric stability (LTS). Our goals are to (i) explain why the mixed scheme outperforms both Tiedtke and pure ML, (ii) identify regimes of high/low model confidence and its spatial structure, and (iii) characterize conditional mean heating and moistening profiles as functions of CWV and LTS. These analyses provide process-level insight into the hybrid model's strengths, demonstrate improved precipitation skill, and clarify how convective processes interact with the large-scale climate as constrained by observational products.

\begin{figure}[tbh]
    \centering
    \includegraphics[width=1\linewidth]{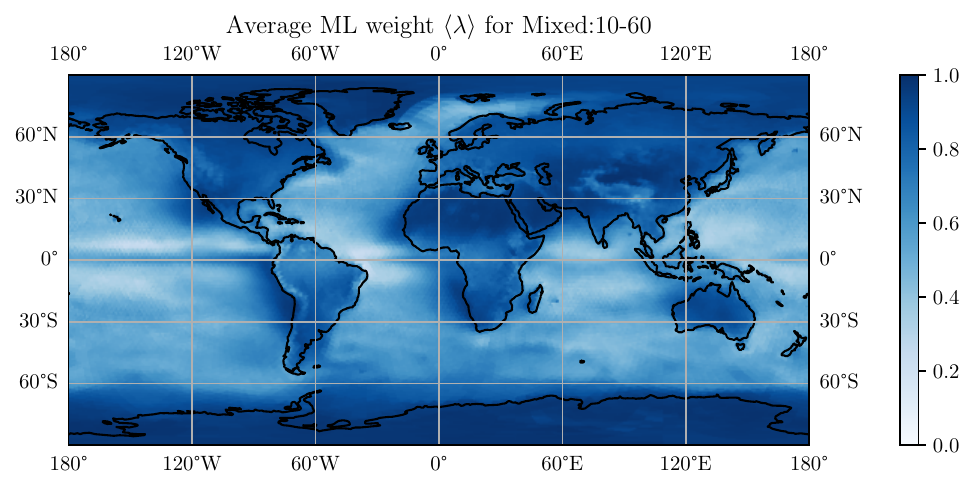}
    \caption{The spatial distribution of the temporally-averaged ML weight $\langle\lambda\rangle$ over one year of simulation for the Mixed:10-60 model with a physics-informed weight $\alpha=0.1$. The overall time averaged ML weight was $\langle\lambda\rangle\approx 0.67$ for the coupled run.}
    \label{fig:mlweight_spatial_distribution}
\end{figure}

\begin{figure}[tbh]
    \centering
    \includegraphics[width=1\linewidth]{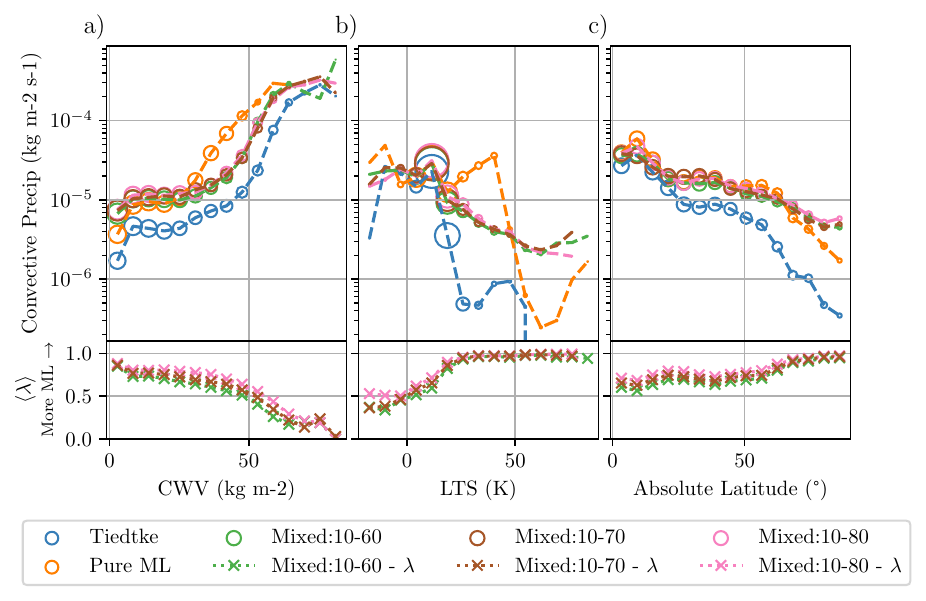}
    \caption{Conditionally averaged convective precipitation (top row) and average ML weight $\langle\lambda\rangle$ (lower row) as a function of CWV (a), lower tropospheric stability (LTS) (b), and absolute latitude (c). Circles represent the convective precipitation (circle sizes indicate the number of samples in the respective region) and crosses the average ML weight $\langle\lambda\rangle$. All plots within one row share the same y-axis scale.}
    \label{fig:convective_precip_cwv_lts_abslat}
\end{figure}

First, we investigate the spatial distribution of the average weight, $\langle\lambda\rangle$, for the Mixed:10-60 model with $\alpha=0.1$ (\Cref{fig:mlweight_spatial_distribution}).
The average ML weight is generally higher over land than over oceans, reflecting greater confidence in ML predictions in continental environments.
Furthermore, the model exhibits increased confidence in high-latitude regions compared to the tropics.
In the tropics, where convective activity is abundant, the model’s confidence is reduced, likely due to inherent variability in this region.
This fits the observation in \Cref{fig:convective_precip_cwv_lts_abslat} that the ML models' confidence decreases with the magnitude of the column water vapor in the column as higher magnitudes of water vapor are expected in the tropics.
Moreover, in the tropics the predicted tendencies and convective precipitation have larger absolute magnitudes, which in turn generate larger prediction errors and reduce the ML model's confidence.
Importantly, regions with complex orography – including the Himalayas, Andes, Ethiopian Highlands, and Rocky Mountains – tend to exhibit lower model confidence, even without explicitly providing orographic information to the ML models.
The overall fractions of precipitation originating from convective (ML), convective (Tiedtke), and large-scale are \qty{62}{\percent}, \qty{5}{\percent}, and \qty{33}{\percent}.
The convective (ML) precipitation fraction therefore outweighs both the Tiedtke precipitation and large-scale precipitation on average and is close to the average ML weight with \qty{67}{\percent}.

For comparison, the spatial distribution of the average ML weight is shown for two more models in \Cref{fig:mlweight_spatial_distribution_appendix}. The patterns are very similar, but the overall ML weight increases with higher $p_1$ values as expected.

To understand under which conditions the ML-based schemes predict convective precipitation, \Cref{fig:convective_precip_cwv_lts_abslat} shows the conditionally averaged convective precipitation and average ML weights $\langle\lambda\rangle$ predicted by different schemes as a function of cumulative CWV and lower tropospheric stability, defined as 

\begin{equation}
    \mathrm{LTS}=\theta_{\qty{\sim 700}{\hecto\pascal}}-T_\mathrm{sfc},
\end{equation}

with the potential temperature $\theta$ at approximately \qty{700}{\hecto\pascal} and the surface temperature $T_\mathrm{sfc}$.
The LTS can be regarded as measuring the strength of the inversion that caps the planetary boundary layer \cite{wood2006RelationshipStratiform,brenowitz_interpreting_2020}.

Panel (a) of \Cref{fig:convective_precip_cwv_lts_abslat} reveals that the curves show comparable behaviors, especially among all mixed models, similarly to panels (b) and (c).
Notably, the mixed models and the Tiedtke show a sharp pickup of precipitation around \qtyrange{50}{60}{\milli\meter} globally, similar to the critical value of \qty{66}{\milli\meter} reported for tropical environments in \citeA{holloway_moisture_2009}.
The Tiedtke scheme robustly shows the lowest precipitation values for all CWV conditions, consistent with \Cref{fig:precipitation_zonal_and_extremes} (b).
In contrast, the pure ML model exhibits relatively low precipitation for low CWV but high precipitation for mid-level CWV values.
For very high CWV values, the schemes show slightly different behavior, although it is notable that this region contains very few samples.
The decreasing ML model confidence (hence increasing $\lambda$) observed as CWV is increased therefore results from both the scarcity of training samples and the large inherent variability associated with convective processes in this region of the CWV space \cite{jones_global_2004,sukovich_extreme_2014,bretherton_relationships_2004}.

In panel (b) of \Cref{fig:convective_precip_cwv_lts_abslat}, the mixed models vary more smoothly with LTS than either Tiedtke or the pure ML models, which show discontinuities. Tiedtke also shuts down convection quickly at high LTS, likely missing cases where large-scale forcing (e.g., mesoscale convective systems or at higher latitudes) can trigger convection under relatively stable conditions. This helps explain why mixed schemes that place more weight on the ML component at high latitudes perform best, e.g., the 10–60 mixed model depicted in \Cref{fig:mlweight_spatial_distribution}. As expected, convective precipitation generally increases with decreasing dry stability (decreasing LTS).
The Tiedtke scheme shows a sudden decrease in precipitation for very low LTS values, although it is worth noting that this region contains very few samples.
The ML weight, i.e. $\langle\lambda\rangle$, of the models initially exhibits a modest increase (or even a slight decrease) as dry stability increases, but then rises more sharply until an LTS of \qty{25}{\kelvin} is reached, after which it levels off and remains almost constant close to 1 under more stable conditions.
This trend is reasonable because convective precipitation is expected to be low under very stable atmospheric conditions and more intense and difficult to predict for unstable environments.
Notably, the mixed models predict higher precipitation rates than both the Tiedtke and the Pure ML model. One reason for this counterintuitive behavior is that the Pure ML model is different from the one used in the mixed models since it is trained with PINN weight $\alpha=0$. Another aspect is that due to the coupling with different schemes, the coupled model (potentially) runs in different dynamical regimes, and we only condition on LTS here without fixing any other variable in the analysis. As a consequence one cannot expect the model to ``interpolate'' between the behavior of the Pure ML model and the Tiedtke scheme for these kinds of emergent statistics.

The convective precipitation decreases with increasing latitude (Panel (c)), as expected. In contrast, the ML weight increases with absolute latitude, reaching values close to 1 for latitudes exceeding \qty{80}{\degree}, consistent with the patterns observed in \Cref{fig:mlweight_spatial_distribution}.
Similarly to Panel (a), the Tiedtke scheme demonstrates the lowest convective precipitation for almost all data points while the pure ML model and also the mixed models, predict relatively high values overall.

Taken together, \Cref{fig:convective_precip_cwv_lts_abslat} illustrates that when the mixed model parameterizations are observationally informed, the resulting schemes predominantly converge toward the behavior of purely data-driven approaches across a wide range of atmospheric conditions.
However, under moist and unstable conditions, the mixed schemes exhibit a modest shift toward the conventional Tiedtke scheme. This calibration enables a more robust interpretation of convective processes by constraining the inverse problem of mapping convective tendencies as a function of column water vapor, lower-tropospheric stability, and geographic context.
The resulting parameterizations yield physically interpretable regime behavior while mitigating the risk of extrapolation in regions of low confidence. 

As illustrated in \Cref{fig:mlweight_spatial_distribution}, the ML weight exhibits a dependence on both latitude and topography.
To further investigate this relationship, \Cref{fig:convective_precip_zsfc} presents the convective precipitation and ML weight as functions of the surface height.
The convective precipitation displays a non-monotonic relationship with surface height, characterized by an initial decrease followed by a sharp increase at high elevations (above \qtyrange{3}{4}{\kilo\meter}). The relatively low (but still over \qty{60}{\percent}) ML weights obtained for sea surface heights are consistent with the challenges associated with predicting convection within the tropics and Intertropical Convergence Zone (ITCZ).
Furthermore, the ML weight decreases moderately at high surface heights, indicating a subtle dependence on topography in these regions.

\begin{figure}[tbh]
    \centering
    \includegraphics[width=1\linewidth]{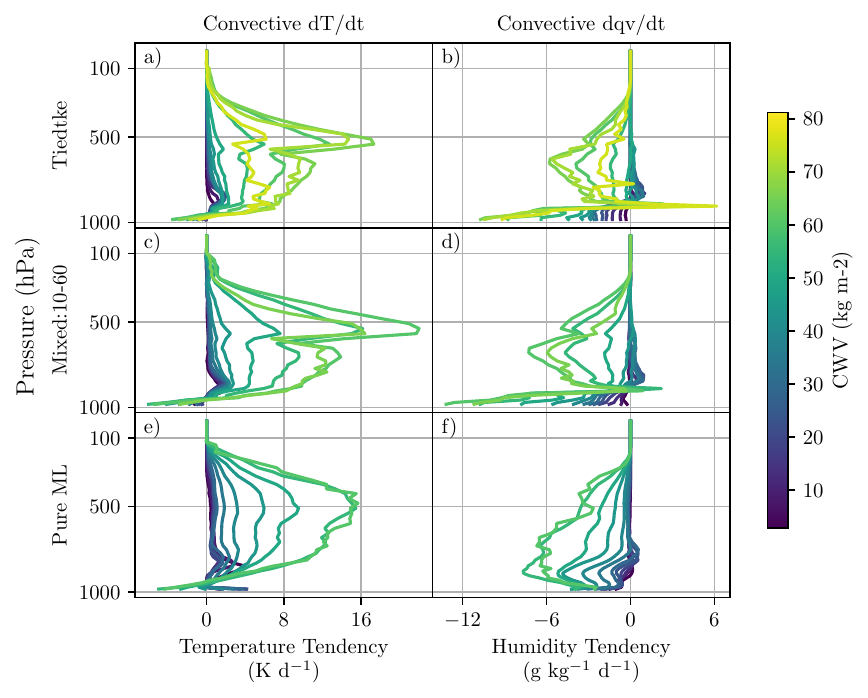}
    \caption{Conditional averages of convective heating rates (first column) and moistening rates (second column) as a function of height. The conditioning is based on CWV while we keep the value for the LTS fixed to $\mathrm{LTS}=\qty{11.4}{\kelvin}$. Each row corresponds to a different coupled scheme: (a,b) for Tiedtke, (c,d) for Mixed:10-60, and (e,f) for the pure ML scheme. Conditional averaged curves are only computed for CWV conditions having at least ten samples.}
          \label{fig:1d_average-profiles_fixedlts}
   \end{figure}

To investigate how the 3D outputs of the ML/mixed scheme behave we now turn our attention to profiles of the convective temperature and humidity tendencies as well as the corresponding enthalpy changes conditionally averaged on CWV for the Mixed:10-60 model with $\alpha=0.1$. These profiles are displayed in \Cref{fig:1d_average-profiles_fixedlts} for different values of CWV.
These correspond to the transects visualized as dashed red lines in \Cref{fig:precipitation_prw_lts_2d}. 
Similar profiles conditionally averaged on LTS are shown in \Cref{fig:1d_average-profiles_fixedcwv}.

A comparison between the ML/mixed schemes and the Tiedtke scheme reveals similarities in the heating rate behavior, as evident in panels (a,c,e). The mixed scheme exhibits slightly higher tropospheric heating rates and correspondingly lower surface heating rates than the Tiedtke scheme. In contrast, the pure ML scheme displays a similar overall magnitude, but with smoother profiles as a function of height. Notably, the ML scheme lacks the mid-tropospheric decrease in heating rates observed at higher humidity values, distinguishing it from the other two schemes. The analysis may exhibit a slight bias towards higher CWV values and a relatively low ML weight, correspondingly (\Cref{fig:convective_precip_cwv_lts_abslat}), due to the x-axis scale. However, by zooming in, the mixed scheme and the Tiedtke schemes still show a high level of similarity.

The moistening rates depicted in panels (b,d,f) show that the mixed scheme closely resembles the Tiedtke scheme, despite the ML weight being approximately $\sim$\qty{67}{\percent} on average.
This suggests that the mixing approach effectively retains the simulation's proximity to the conventional ICON model's distribution, while incorporating ML predictions to enhance agreement with observational data, as evident in \Cref{fig:online_pareto_plots,fig:precipitation_zonal_and_extremes}.
In contrast, the pure ML model yields smoother predictions that lack some features, such as the moistening peak at around \qty{900}{\hecto\pascal}.

It is worth noting that for the shown profiles, the mixed model predicts heating, moistening, and precipitation in a manner that nearly conserves enthalpy, whereas the pure ML model exhibits net fluxes into the column of up to \qty{50}{\watt\per\metre^2}, indicating a notable deviation from enthalpy conservation as already seen in \Cref{fig:enthalpy_residuals_online}.
The mean absolute enthalpy residuals are \qty{0.003}{\watt\per\meter^2}, \qty{1.024}{\watt\per\meter^2}, \qty{26.037}{\watt\per\meter^2} for the Tiedtke, Mixed:10-60, and pure ML scheme, respectively.
The residual of the pure ML model is therefore higher than for the Mixed:10-60 model by factor of over 25.
Looking at the ML weight $\langle\lambda\rangle$, conditionally averaged for the same conditions, we find that the weight has an average magnitude of $\langle\lambda\rangle\approx 0.65$.
This showcases that the reduced enthalpy residual is not only due to mixing with the Tiedtke scheme but also to introducing the physics-informed loss terms (see \Cref{eq:e_residual,eq:m_residual,eq:u_residual,eq:v_residual}) during training.

For the tendencies and enthalpy changes for varying LTS and fixed \qty{19.6}{\kilogram\per\meter^2} displayed in \Cref{fig:1d_average-profiles_fixedcwv}, the profile comparison is less clear since the Tiedtke scheme shows a high variability, especially for lower layers.
In general, the mixed model exhibits the smoothest profiles with, e.g., upward moisture transport being more visible than for the Tiedtke scheme.
The net column enthalpy flux reveals the same behavior as the pure ML scheme is far from conserving enthalpy, while the mixed scheme is much closer to conservation.

\subsection{Twenty-year AMIP run}\label{sec:results:20_year_amip}

In this section, we evaluate AMIP-style simulation runs for 20 years (1979-1998) with the presented ML and mixed schemes.
We have already demonstrated the stability and skill of the method for one year long simulations, but longer simulations remain to be investigated.

Online runs with the originally developed schemes became numerically unstable after 1.5 - 3 years. As the schemes are trained on the \climsim{} dataset and even under the assumption that they are unbiased estimators of the true subgrid tendencies on this dataset, the transfer to the new domain (ICON) can transform them into biased estimators.
Therefore, small errors can add up over time and finally lead to the coupled model diverging.

Using the method introduced in \Cref{sec:methods:noise}, we therefore made the schemes more robust by dynamically adjusting the noise variance such that the model maximally loses $\Delta R^2$ of its predictive skill while increasing its robustness through the addition of noise. We applied this method to the pure ML model and the ML model with a physics-informed weight $\alpha=0.1$ with $\Delta R^2=0.2$. Depending on the chosen initial noise level $\sigma_0$, this process typically takes about \num{5}--\num{10} epochs to converge.

\begin{figure}[tbh]
    \centering
    \includegraphics[width=1\linewidth]{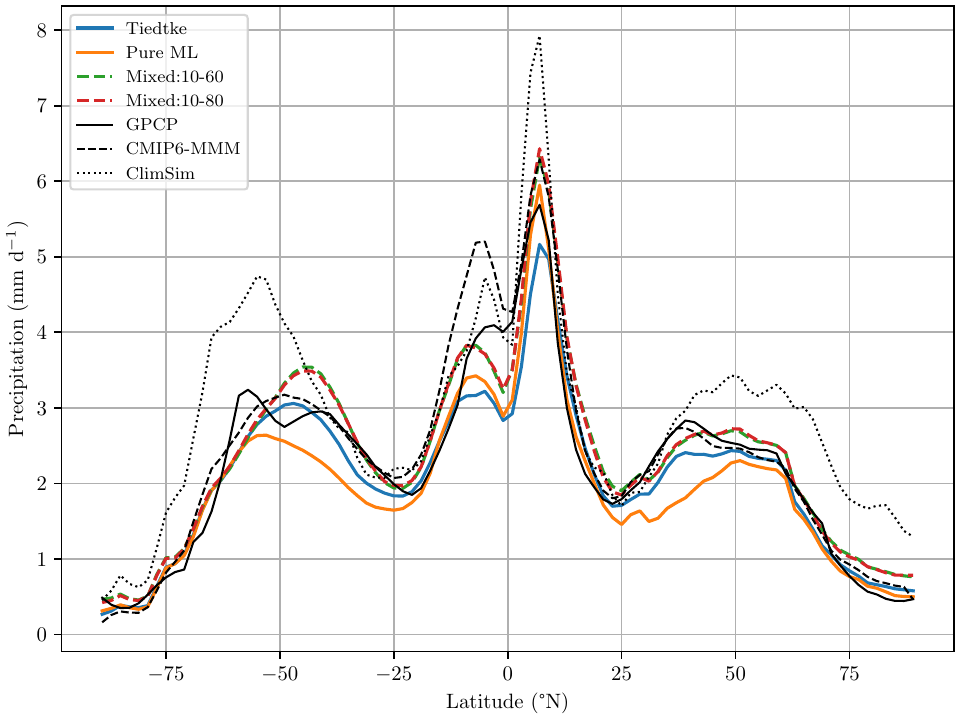}
    \caption{Zonal mean precipitation evaluated over twenty years for the observational dataset (GPCP), the Tiedtke scheme, the pure ML scheme, the Mixed:10-60 scheme, the Mixed:10-80 scheme, the CMIP6 MMM, and the \climsim{} dataset. For \climsim{}, the zonal mean precipitation is evaluated over its available 10-year simulation period.}
    \label{fig:zonal_mean_precip_20y}
\end{figure}

\begin{table}[htb]
    \centering
    \begin{tabular}{l l l l l l}
        \toprule
        Data & $A_P$ & $E_P$ & $A_P$ Bias & $E_P$ Bias & RMSE (\unit{{\milli\meter\per\day}})\\
        \midrule
        GPCP & 0.189 & 1.163 & - & - & - \\
        \hline
               Tiedtke & 0.247 & 0.909 & \textbf{0.058} & -0.254 & 0.382\\
        Pure ML & 0.257 & 0.924 & 0.068 & \textbf{-0.239} & 0.459\\
        Mixed:10-60 & 0.275 & 0.901 & 0.086 & -0.262 & 0.380\\
        Mixed:10-80 & 0.279 & 0.902 & 0.090 & -0.261 & \textbf{0.375}\\
        \hline
        \climsim{} & 0.268 & 0.973 & 0.079 & -0.190 & 0.904\\
        CMIP6-MMM & 0.060 & 1.037 & -0.129 & -0.126 & 0.394\\
                      \bottomrule
    \end{tabular}
    \caption{The tropical precipitation asymmetry index $A_P$ and the equatorial precipitation index $E_P$, and their biases with respect to GPCP for the data shown in \Cref{fig:zonal_mean_precip_20y}.}
    \label{tab:itcz_indices_20_year}
\end{table}

Although \climsim{} shows the smallest deviations in the $A_P$ and $E_P$ metrics, \Cref{fig:zonal_mean_precip_20y} reveals that its zonal distribution deviates substantially from observations.
Notably, precipitation is overestimated in the extratropics, along the ITCZ, and at high latitudes in the Northern Hemisphere.
This larger mean bias is also reflected in the RMSE score for \climsim{} in \Cref{tab:itcz_indices_20_year}, which is approximately twice as high as that of the second-worst model, indicating a significant overall bias.

The CMIP6 multi-model mean (MMM) shows a reasonable zonal mean precipitation distribution in general but has a substantial double ITCZ bias which is also reflected in the highest overall bias of the tropical precipitation asymmetry index $A_P$ of $-0.129$ as reflected in \Cref{tab:itcz_indices_20_year}. The MMM also has a relatively low RMSE with \qty{0.394}{\milli\meter\per\day} but is outperformed by, e.g., the Mixed:10-80 model with a RMSE of \qty{0.375}{\milli\meter\per\day}.

The zonal mean precipitation shown in \Cref{fig:zonal_mean_precip_20y} and the corresponding biases summarized in \Cref{tab:itcz_indices_20_year} indicate that all models produce reasonably realistic distributions over the 20-year simulation period.
Among these models, the Tiedtke model exhibits the smallest bias in the asymmetric precipitation component, while the pure ML model performs best in capturing the symmetric component among the schemes.
The Mixed:10-80 model achieves the lowest RMSE when compared to observational data, despite the relatively high RMSE of the \climsim{} distribution.
The RMSE, which is arguably the more meaningful metric as mentioned in \Cref{sec:results:benchmarking_with_obs}, is slightly better for the mixed model compared to the Tiedtke model with a difference of \qty{0.007}{\milli\meter\per\day}.

\begin{figure}[tbh]
    \centering
    \includegraphics[width=0.96\linewidth]{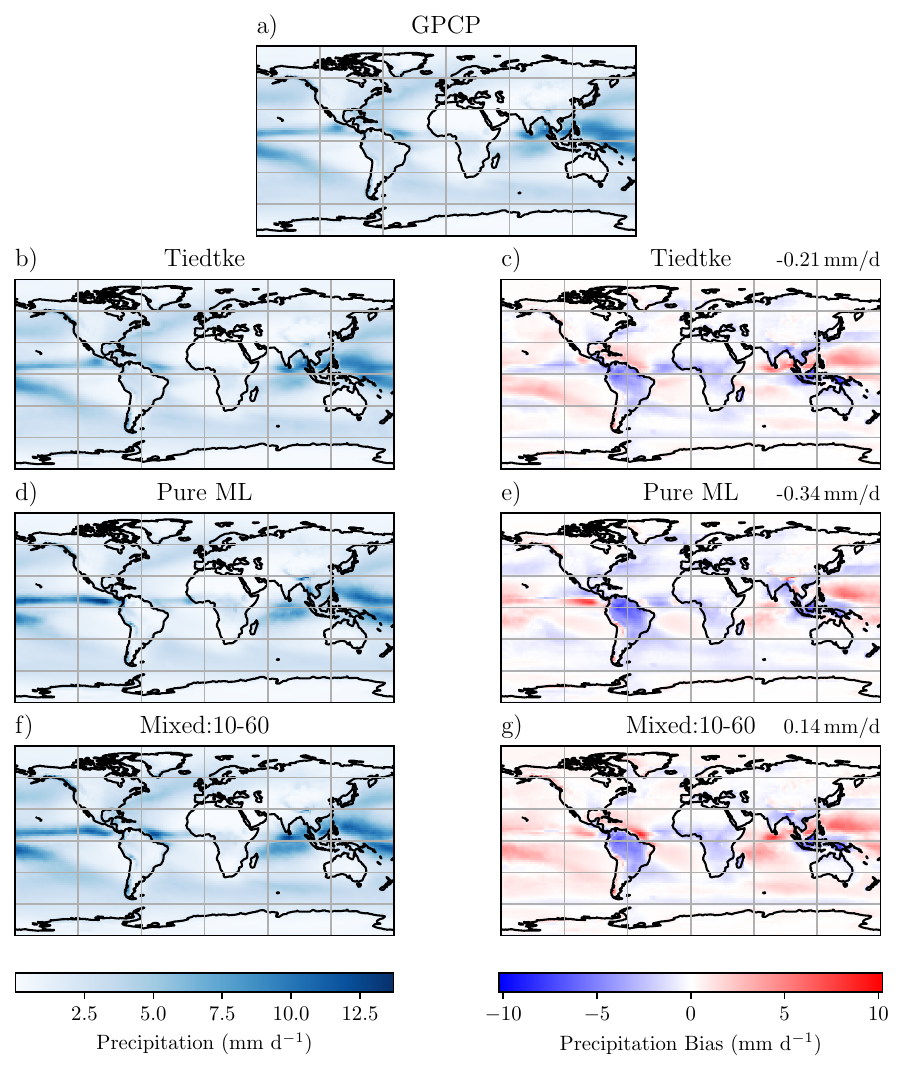}
    \caption{The spatial distribution of 20-year averaged precipitation for different convection schemes in the left column and the bias with respect to GPCP in the right column. The first row (a) shows precipitation for the GPCP data, the Tiedtke scheme in the second row (b-c), the pure ML scheme in the third row (d-e), and the Mixed:10-60 scheme in the last row (f-g). In the upper right of each bias plot, the area-weighted mean bias is displayed.}
    \label{fig:20y_precip_vs_gpcp}
\end{figure}

For the spatial distribution of the mean precipitation shown in \Cref{fig:20y_precip_vs_gpcp}, the Tiedtke and pure ML models show negative biases of \qty{-0.21}{\milli\meter\per\day} and \qty{-0.34}{\milli\meter\per\day}, respectively, whereas the Mixed:10-60 scheme yields a slightly smaller positive bias of \qty{0.14}{\milli\meter\per\day}.
Spatially, the mean biases shown in \Cref{fig:20y_tas_vs_era5} show a very similar distribution with a general slight overestimation of mean precipitation and underestimation patterns mainly seen over low-latitude, continental regions.
Similarly, in terms of near-surface temperature $T_{2m}$ (\Cref{fig:20y_tas_vs_era5}), the Mixed:10–60 model exhibits the smallest mean bias (\qty{-0.26}{\kelvin}) over the 20-year period, compared to the Tiedtke scheme (\qty{0.5}{\kelvin}) and the pure ML model (\qty{1.03}{\kelvin}).
When looking at the timeseries of the global mean near-surface temperature no considerable drift could be observed for all the shown simulations here.
However, while the mean biases highlight the potential of the confidence-guided-mixing approach, the broader set of metrics summarized in \Cref{tab:spatial_mean_bias_20_year} presents a more nuanced picture.
Notably, when evaluating the RMSE and MAE, the Tiedtke scheme still outperforms the mixed approach in some cases.
Overall, these results indicate that while the mixing approach shows promise for long-term climate integrations, its performance enhancements vary depending on the specific metric evaluated.

\section{Summary}\label{sec:conclusion}

Through our proposed confidence-guided mixing, we developed robust parameterizations that yielded successful decade-long runs. Impressively, this is true despite our parameterizations being trained on a dataset generated by another GCM, enabling the ICON-A model to benefit from the advantages of a superparamerized GCM. This study provides a proof-of-concept demonstrating that, through careful data preprocessing and deliberate model design choices — including confidence-guided mixing, loss function design, physics-informed training, and additive noise injection — it is possible to transfer ML convection schemes from one GCM to another without compromising stability and accuracy. The mean weight given to the ML-transferred parameterization is $\approx 0.67$, confirming a fundamental change in the convective parameterization’s behavior rather than a simple bias correction of the Tiedtke scheme.

When training on the \climsim{} dataset, we first separated the radiative from the convective heating tendencies using the \radscheme{} radiation scheme. To achieve this, we modified the scheme slightly to match the version used in ICON and to allow us to input full columns from the \climsim{} data as explained in \Cref{sec:data:radremoval}. We note that this separation represents an approximation of the true radiative tendencies employed by the E3SM-MMF model, as the radiation scheme was run for multiple radiation columns in each grid cell of the multiscale modeling framework, and we only have access to the coarse-grained state in \climsim{}. Future versions of \climsim{} would benefit from outputting radiative tendencies explicitly, enabling process-based training rather than emulating all subgrid physics. Likewise, the SRMs in E3SM still parameterize sub-SRM processes (e.g., turbulence, microphysics), which contribute to \climsim{} tendencies; outputting those terms separately would further facilitate process-based schemes. The training would also benefit from a more accurate representation of precipitation in the \climsim{} data (see \Cref{fig:zonal_mean_precip_20y}).

The inclusion of physics-informed terms in the loss function improves model performance across various metrics. Specifically, adding the residuals of conserved quantities to the loss function led to improved conservation online, as evident in \Cref{fig:enthalpy_residuals_online}. However, it is likely that using a training dataset where conservation laws can be strictly enforced without any net in- or out-fluxes into the columns would further improve the method. Creating such a dataset would be a crucial next step in further improving the here shown proof-of-concept method.

After generating the training data and designing the model and loss function, we performed a thorough hyperparameter search, an essential step for finding a good trade-off between accuracy and computational efficiency, with the number of multiply-accumulate operations proving to be a well-suited measure of computational complexity (for CPU inference). Our results revealed 181 candidate schemes along the Pareto front when comparing different metrics. Some of these models were found to perform even better than the conventional Tiedtke parameterization used in ICON, a promising outcome considering that \icon{} has been calibrated to behave optimally with the Tiedtke scheme. In particular, the representation of precipitation, water vapor, and near-surface temperature potentially benefits from the confidence-guided mixing approach as demonstrated in \Cref{fig:online_pareto_plots}.

Investigating the conditions under which the ML/mixed schemes produce convective precipitation revealed a reasonable behavior, with precipitation generally increasing with higher column water vapor and decreasing with higher atmospheric dry stability as shown in \Cref{fig:convective_precip_cwv_lts_abslat}. Notably, the mixed scheme does not fully shut down convection under high dry stability conditions, which may help when convection is forced by, e.g., large-scale horizontal advection or orographic forcing.
Moreover, we observed that the confidence of the mixed schemes decreased in regimes with few training samples as well as in regions characterized by high variability of precipitation. Conditionally averaged heating and moistening profiles in \Cref{fig:1d_average-profiles_fixedlts} show substantial differences between the pure-ML, mixed, and Tiedtke schemes. Despite an average ML contribution of approximately \qty{\sim 67}{\percent}, the mixed scheme resembles the conventional ICON model in its physical behavior more closely than the pure ML model, maintaining dynamical consistency and avoiding out-of-distribution predictions while still leveraging the ML component’s learned physical relationships. Additionally, our analysis of the enthalpy profiles demonstrated again that the mixed scheme learned with a physics-informed weight of only 0.1 substantially improved conservation of enthalpy.
These results, and the chosen tuning parameters for further analysis, were based on one-year-long simulations and cannot be expected to be robust due to the short evaluation period. However, as a proof-of-concept, they show that the schemes could be adjusted to work well, even outperforming Tiedtke for some metrics. By analyzing their emergent precipitation statistics, we additionally showed that they can potentially be tuned to observations and learned from. Performing 20-year-long simulations for all 181 candidate schemes would have been computationally infeasible.

Finally, as demonstrated in \Cref{sec:results:20_year_amip}, we achieved long-term stability using an engression-like technique, which provided data-driven extrapolation by effectively forcing the ML model to behave smoothly for small input perturbations. This result could potentially help many more ML-based parameterization schemes which very commonly struggle with long-term stability when coupled to GCMs. The results regarding precipitation and temperature patterns shown in \Cref{sec:results:benchmarking_with_obs,sec:results:20_year_amip} indicate that the pure ML and mixed schemes are capable of generating realistic patterns, which for near-surface temperature even outperform the Tiedtke baseline with respect to observational references by having a mean bias about half as large as for the Tiedtke model for the 20-year evaluation as shown in \Cref{tab:spatial_mean_bias_20_year}. However, calibration against observational data may further enhance the predictive skill of all models examined.

As illustrated in \Cref{fig:mlweight_spatial_distribution} and also \Cref{fig:convective_precip_cwv_lts_abslat}, the ML scheme exhibits relatively high confidence in the extratropics and high latitudes while maintaining a non-zero contribution in the tropical regions that were used to design the Tiedtke scheme. The examination of the results from the twenty-year-long simulations in \Cref{fig:zonal_mean_precip_20y} suggests that this confidence may be overestimated due to out-of-distributions estimates, highlighting the potential benefit of developing a separate convective triggering scheme or lowering the scheme's confidence (i.e. the $p_0$ and $p_1$ parameters) to improve overall model performance.
Moreover, training on a dataset which is closer to observational references for, e.g., the zonal mean precipitation (\Cref{fig:zonal_mean_precip_20y}) would also benefit the model developement.

As we developed a tunable ML-based scheme, future work should also prioritize proper tuning, exploring various settings of parameters such as $p_0$, $p_1$, the level of stochastic noise injection, and the weighting $\alpha$ of physical loss terms in the hybrid objective function to further optimize the scheme's performance. Furthermore, the confidence estimates produced by the ML model could be leveraged to develop a stochastic parameterization framework, transforming the current deterministic predictions into probabilistic outputs. Such a stochastic formulation would better represent subgrid variability and improve the representation of uncertainty in climate and weather simulations. Another direction for future research could be reducing the computational complexity (see \Cref{fig:offline_HPO_pareto}) of the used schemes even more since the coupled \icon{} runs are roughly half as fast as the runs with the Tiedtke scheme.

Ultimately, our goal is to implement an ML-based convection scheme into ICON-XPP-MLe (where XPP stands for eXtended Predictions and Projections and MLe for machine learning enhanced) \cite{egusphere-2025-2473}. Realizing this goal will require further work before the current proof-of-concept can be effectively deployed within this hybrid ESM. This will include systematic tuning of the scheme and hybrid ESM, potentially through automated methods such as the approach proposed by \citeA{grundner_reduced_2025}, further testing, and potentially interpolating the training data to the vertical levels of ICON-XPP-MLe. This would ensure seamless integration and optimal performance of the ML-based parameterization scheme within the broader modeling framework. Another important direction for future research is to assess the sensitivity of the ML scheme to horizontal resolution.
We plan to evaluate its performance at higher resolutions, such as \qty{80}{\kilo\meter} × \qty{80}{\kilo\meter}, to determine its scalability and robustness across different model configurations.
This will help clarify whether the learned relationships generalize across resolutions or require designing a scale-aware version of the scheme.

Additionally, a direct integration with the ICON-XPP-MLe modeling framework may be facilitated by incorporating ICON-specific simulation data into the training pipeline. A suitable dataset would have to fulfill several constraints regarding the length of the simulated time period and spatial extent, frequency of output, as well as a convenient (though not strictly necessary) scale separation, as mentioned in the introduction. Given such a dataset, the inclusion of \icon{} data may be achieved either through retraining the model on the ICON output or by applying transfer learning techniques to adapt the existing models further to the \icon{} model.

Together, these developments, ranging from stochastic extensions to resolution dependence studies and model-specific adaptation, will be crucial for advancing the reliability, robustness, and applicability of ML-based parameterizations in long-term climate simulations.

\clearpage
\appendix
\section{Appendix}

\subsection{Non-dimensionalization of Residual Fluxes}\label{sec:app:non-dim}

As written in \Cref{sec:method:loss_func:physicsloss}, we start with the chosen scaling constants as they are defined in \icon{} (except $t_0$):

\begin{align*}
    g_0&=\qty{9.80665}{\metre\per\s\tothe{2}},\\
    t_0&=\qty{10}{\s},\\
    \rho_\mathrm{h2o}&=\qty{1000}{\kg\per\metre\tothe{3}},\\
    c_{p}&=\qty{1004.64}{\joule\per\kelvin\per\kg}.
\end{align*}

We use these constants to derive scales for length $l_0$, temperature $T_0$, energy density $e_0$, mass flux $m_0$, velocity $v_0$, and pressure $p_0$:

\begin{equation*}
    l_0=g_0\,t_0^2,\quad
    T_0=\frac{e_0}{c_p},\quad
    e_0=g_0^2\,t_0^2,\quad
    m_0 = \rho_\mathrm{h2o}\,g_0\,t_0,\quad
    v_0 = g_0\,t_0,\quad
    p_0=\rho_\mathrm{h2o}\,g_0^2\,t_0^2.
\end{equation*}

Furthermore, the latent heat of vaporization $L_v = \qty{2.5008e6}{\joule\per\kilogram}$ and sublimation $L_s = \qty{2.8345e6}{\joule\per\kilogram}$ are non-dimensionalized by dividing by $e_0$.

In ICON, the net column in \& out fluxes for enthalpy, mass, zonal, and meridional momentum can be formulated as follows:

\begin{align}
    H_\mathrm{res} &= \int_{z_\mathrm{bot}}^{z_\mathrm{top}} \rho\,\left(\frac{\partial T}{\partial t}\,c_p - \frac{\partial q_\mathrm{l}}{\partial t}\,L_\mathrm{v} - \frac{\partial q_\mathrm{i}}{\partial t}\,L_\mathrm{s}\right) dz - L_\mathrm{v}\,{\cal P}_{\mathrm{rain}}- L_\mathrm{s}\,{\cal P}_{\mathrm{snow}}, \\
    m_\mathrm{res} &= \int_{z_\mathrm{bot}}^{z_\mathrm{top}}  \rho\,\left(\frac{\partial q_\mathrm{v}}{\partial t} + \frac{\partial q_\mathrm{l}}{\partial t} + \frac{\partial q_\mathrm{i}}{\partial t}\right) dz + {\cal P}_{\mathrm{rain}}+ {\cal P}_{\mathrm{snow}}, \\
    u_\mathrm{res} &= \int_{z_\mathrm{bot}}^{z_\mathrm{top}} \rho\,\frac{\partial u}{\partial t}\ dz, \\
    v_\mathrm{res} &= \int_{z_\mathrm{bot}}^{z_\mathrm{top}} \rho\,\frac{\partial v}{\partial t}\ dz.
\end{align}

Using hydrostatic equilibrium for the background vertical coordinate: 

\begin{equation}
    dp = -\rho\,g_0\,dz,
\end{equation}

we convert the vertical integration coordinate from elevation to pressure:

\begin{align}
    H_\mathrm{res} &= \int_{p_\mathrm{top}}^{p_\mathrm{bot}} \frac{1}{g_0}\,\left(\frac{\partial T}{\partial t}\,c_p - \frac{\partial q_\mathrm{l}}{\partial t}\,L_\mathrm{v} - \frac{\partial q_\mathrm{i}}{\partial t}\,L_\mathrm{s}\right) dp - L_\mathrm{v}\,{\cal P}_{\mathrm{rain}}- L_\mathrm{s}\,{\cal P}_{\mathrm{snow}}, \\
    m_\mathrm{res} &= \int_{p_\mathrm{top}}^{p_\mathrm{bot}} \frac{1}{g_0}\,\left(\frac{\partial q_\mathrm{v}}{\partial t} + \frac{\partial q_\mathrm{l}}{\partial t} + \frac{\partial q_\mathrm{i}}{\partial t}\right) dp + {\cal P}_{\mathrm{rain}}+ {\cal P}_{\mathrm{snow}}, \\
    u_\mathrm{res} &= \int_{p_\mathrm{top}}^{p_\mathrm{bot}} \frac{1}{g_0}\frac{\partial u}{\partial t}\ dp, \\
    v_\mathrm{res} &= \int_{p_\mathrm{top}}^{p_\mathrm{bot}} \frac{1}{g_0}\frac{\partial v}{\partial t}\ dp.
\end{align}

Finally, substituting all dimensional quantities with their respective non-dimensional counterparts (marked by a tilde) times the corresponding physical scale yields the following non-dimensional fluxes of enthalpy, mass, zonal and meridional momentum as shown in \Cref{sec:method:loss_func:physicsloss}:

\begin{align}
    \widetilde{H}_\mathrm{res} &= \int_{\tilde{p}_\mathrm{top}}^{\tilde{p}_\mathrm{bot}} \left(\frac{\partial \widetilde{T}}{\partial \widetilde{t}} - \frac{\partial \widetilde{q_\mathrm{l}}}{\partial \widetilde{t}}\cdot \widetilde{L}_\mathrm{v} - \frac{\partial \widetilde{q_\mathrm{i}}}{\partial \widetilde{t}}\cdot \widetilde{L}_\mathrm{s}\right) d\tilde{p} - \widetilde{L}_\mathrm{v}\cdot \widetilde{\cal P}_{\mathrm{rain}} - \widetilde{L}_\mathrm{s}\cdot \widetilde{\cal P}_{\mathrm{snow}}, \\
    \widetilde{m}_\mathrm{res} &= \int_{\tilde{p}_\mathrm{top}}^{\tilde{p}_\mathrm{bot}}  \left(\frac{\partial \widetilde{q_\mathrm{v}}}{\partial \widetilde{t}} + \frac{\partial \widetilde{q_\mathrm{l}}}{\partial \widetilde{t}} + \frac{\partial \widetilde{q_\mathrm{i}}}{\partial \widetilde{t}}\right) d\tilde{p} + \widetilde{\cal P}_{\mathrm{rain}} + \widetilde{\cal P}_{\mathrm{snow}}, \\
    \widetilde{u}_\mathrm{res} &= \int_{\tilde{p}_\mathrm{top}}^{\tilde{p}_\mathrm{bot}}  \frac{\partial \widetilde{u}}{\partial \widetilde{t}}\ d\tilde{p}, \\
    \widetilde{v}_\mathrm{res} &= \int_{\tilde{p}_\mathrm{top}}^{\tilde{p}_\mathrm{bot}}  \frac{\partial \widetilde{v}}{\partial \widetilde{t}}\ d\tilde{p}.
\end{align}

\clearpage

\subsection{The Hyperparameter Optimization Search Space and Offline $R^2$ Scores}
\label{sec:app:opt}

\begin{table}[htb]
    \centering
    \begin{tabular}{l l@{}l@{}l l}
               \toprule
        Parameter & \multicolumn{3}{c}{Search space} & Used in ''Trade-off`` \\
               \midrule
        \texttt{encode\_dim} $e$ & $\{10k $&$\mid k\in\mathbb{N}$,\hspace{1em} &$1\leq k\leq 40\}$ & 280 \\
        \texttt{hidden\_dim} $h$ & $\{10k $&$\mid k\in\mathbb{N}$,&$1 \leq k \leq 40\}$ & 60 \\
        \texttt{iter\_dim} $it$ & $\{100+10 k $\,&$\mid k\in\mathbb{N}$,&$0 \leq k \leq 80\}$ & 120 \\
        \texttt{lstm\_layers} & $\{k $&$\mid k\in\mathbb{N}$,&$1 \leq k \leq 10\}$ & 4\\
        \texttt{dropout\_rate} & \multicolumn{3}{l}{\makecell[cl]{\{0,0.01,0.02,0.03,0.05,0.07,0.1,0.13,\\0.16,0.2,0.25,0.3\}}} & 0.02 \\
        \texttt{learning\_rate} & \multicolumn{3}{l}{\{\num{1e-3},\num{5e-3},\num{6.5e-3},\num{1e-4}\}}  & \num{1e-3} \\
        \texttt{weight\_decay} &  \multicolumn{3}{l}{\{\num{2e-4},\num{1e-2}\}}  & \num{1e-2} \\
        \texttt{batch\_dim} $b$ & \multicolumn{3}{l}{\{256,512,1024,2048\}} & 256 \\
        \texttt{scheduler} & \multicolumn{3}{l}{\{None,\texttt{consanh},\texttt{reduce\_plat}\}} & None \\
        \texttt{optimizer} & \multicolumn{3}{l}{$\varnothing$} & \texttt{AdamW} \\
        \texttt{early\_stopping\_patience} & \multicolumn{3}{l}{$\varnothing$} & 6 \\
        \texttt{input\_dim} $i$ & \multicolumn{3}{l}{$\varnothing$} & 17 \\
        \texttt{column\_height} $l$ & \multicolumn{3}{l}{$\varnothing$} & 42 \\
        \texttt{scalar\_out\_dim} $s$ & \multicolumn{3}{l}{$\varnothing$} & 6 \\
        \texttt{profile\_out\_dim} $p$ & \multicolumn{3}{l}{$\varnothing$} & 2 \\
               \bottomrule
    \end{tabular}
    \caption{The parameter search space used for creating \Cref{fig:offline_HPO_pareto} and the parameter setting for the ''Trade-off`` model. Additionally, some fixed Hyperparameters are indicated with an empty set as the search set. The scheduler \texttt{cosanh} is short for the PyTorch class \texttt{CosineAnnealingWarmRestarts} and \texttt{reduce\_plat} for the class \texttt{ReduceLROnPlateau}. The \texttt{encode\_dim} $e$, \texttt{hidden\_dim} $h$, \texttt{iter\_dim} $it$, \texttt{batch\_dim} $b$, \texttt{input\_dim} $i$, \texttt{column\_height} $l$, \texttt{scalar\_out\_dim} $s$, and \texttt{profile\_out\_dim} $p$ correspond to the dimensions displayed in \Cref{fig:BilstmVis}.}
    \label{tab:HPOs}
\end{table}

\begin{table}[htb]
    \centering
    \begin{tabular}{l l}
        \toprule
        $\alpha$ & offline $R^2$\\
        \midrule
        0 & 0.896\\
        0.01 & 0.894\\
        0.1 & 0.892\\
        0.5 & 0.884\\
        0.9 & 0.631\\
        \bottomrule
    \end{tabular}
    \caption{The overall $R^2$ scores for five models with different weighting factors of the physics informed loss terms.}
    \label{tab:offlineR2}
\end{table}

\clearpage
\subsection{Additional Figures}\label{sec:app:additional_figures}

\begin{figure}[tbh]
    \centering
    \includegraphics[width=1\linewidth]{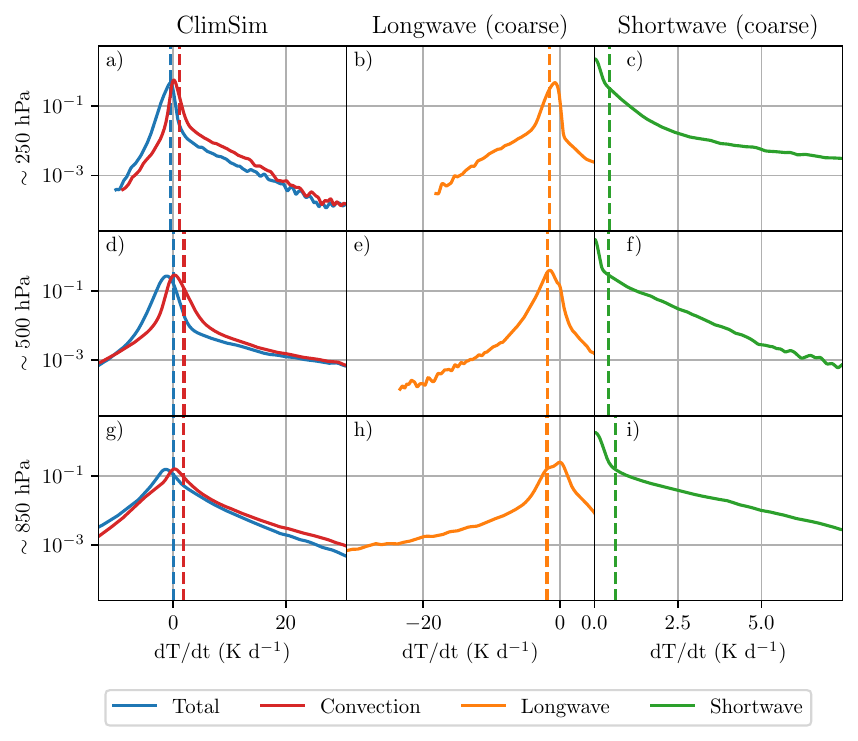}
    \caption{For three pressure levels (rows): (a) temperature tendency distributions before (blue, labeled ``Total'') and after (red, labeled ``Convection'') subtraction of the tendencies computed with \radscheme{}. These radiative tendencies are decomposed into (b) longwave and (c) shortwave components. Mean values are shown with dashed vertical lines for all distributions.}
    \label{fig:rterrtmgp_dTdt_distributions}
\end{figure}

\begin{figure}[tbh]
    \centering
    \includegraphics[width=1\linewidth]{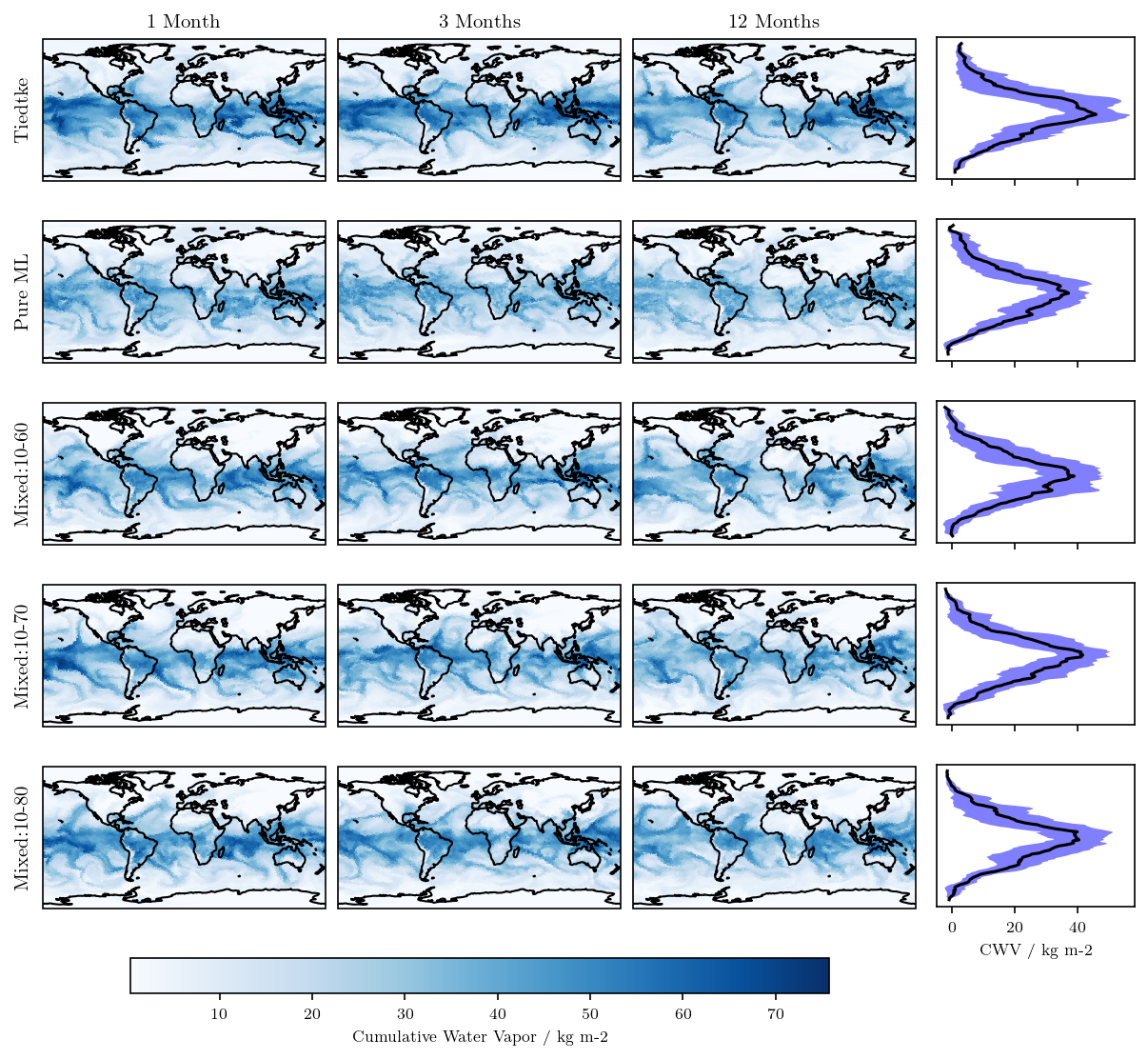}
    \caption{The column water vapor for three simulation snapshots after 1 month (first column), 3 months (second column), and 12 months (third column) of integration. The rows correspond to the five different coupled schemes. The last column shows the zonal mean and standard deviation of the CWV for the last shown timestep of every configuration. The y‐axis corresponds here to the latitudes of the corresponding row.}
    \label{fig:cwv_monthly_evolution}
\end{figure}

\begin{figure}[tbh]
  \centering
  \begin{subfigure}[b]{1\textwidth}
    \begin{overpic}[width=\textwidth]{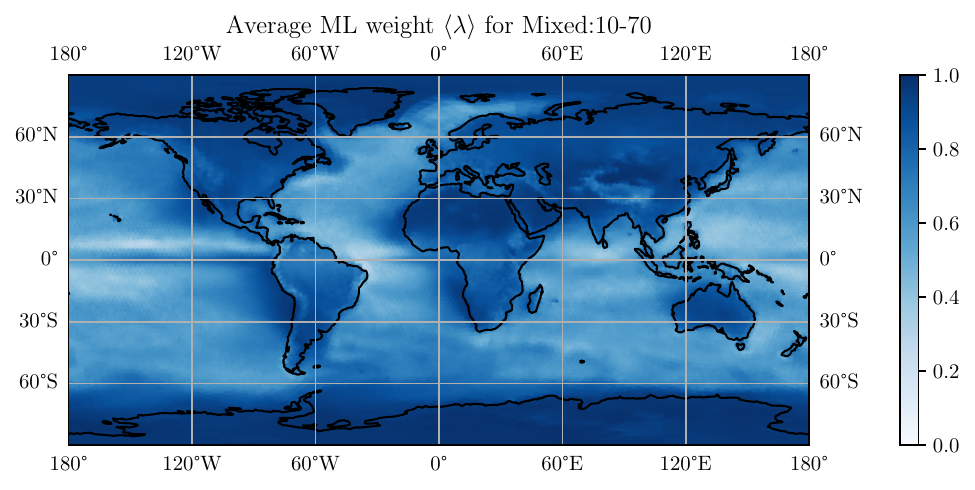}
      \put(0,50){a)}
    \end{overpic}
  \end{subfigure}
  \hfill
  \begin{subfigure}[b]{1\textwidth}
    \begin{overpic}[width=\textwidth]{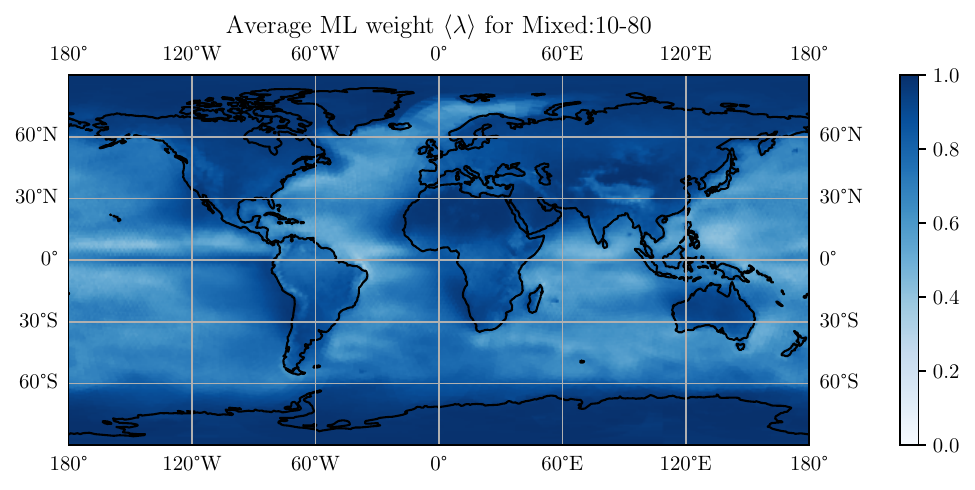}
      \put(0,50){b)}
    \end{overpic}
  \end{subfigure}
  \caption{The spatial distribution of the temporal average ML weight $\langle\lambda\rangle$ over one year of simulation for the Mixed:10-70 and Mixed:10-80 models with a physics-informed weight $\alpha=0.1$. The overall time averaged ML weights were $\langle\lambda\rangle\approx 0.71$ and $\langle\lambda\rangle\approx 0.76$, respectively.}
  \label{fig:mlweight_spatial_distribution_appendix}
\end{figure}

\begin{figure}[tbh]
    \centering
    \includegraphics[width=1\linewidth]{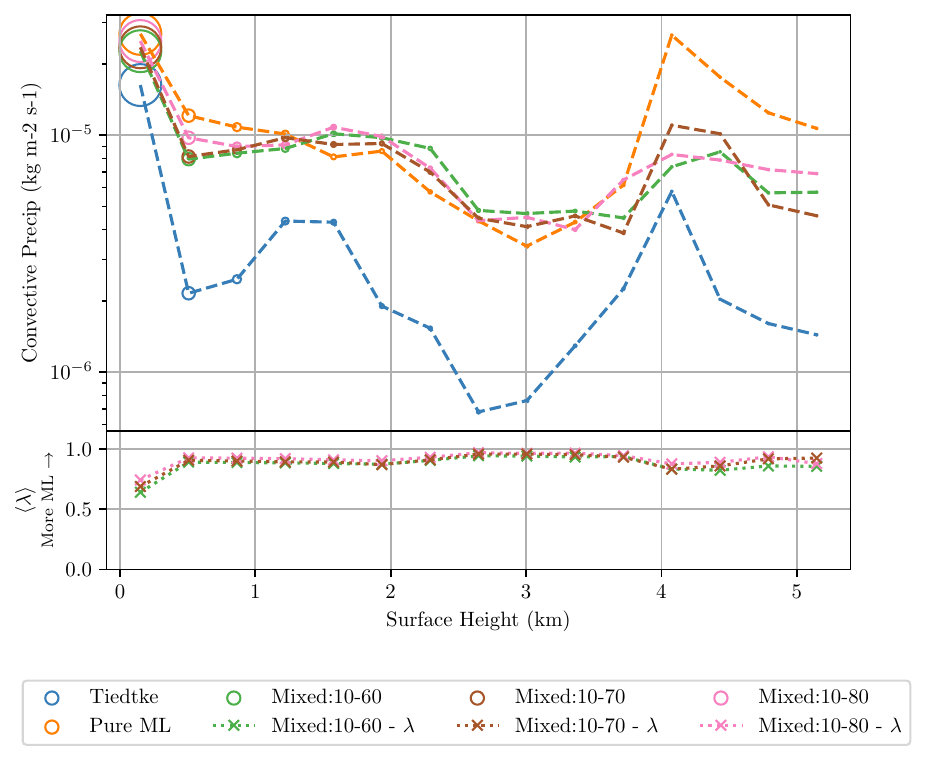}
    \caption{Conditionally averaged convective precipitation as a function of the surface height. Circles represent the convective precipitation (circle sizes indicate the number of samples in the respective region). Crosses in the lower plot represent the average ML weight $\langle\lambda\rangle$.}
    \label{fig:convective_precip_zsfc}
\end{figure}

\begin{figure}[tbh]
    \centering
    \includegraphics[width=1\linewidth]{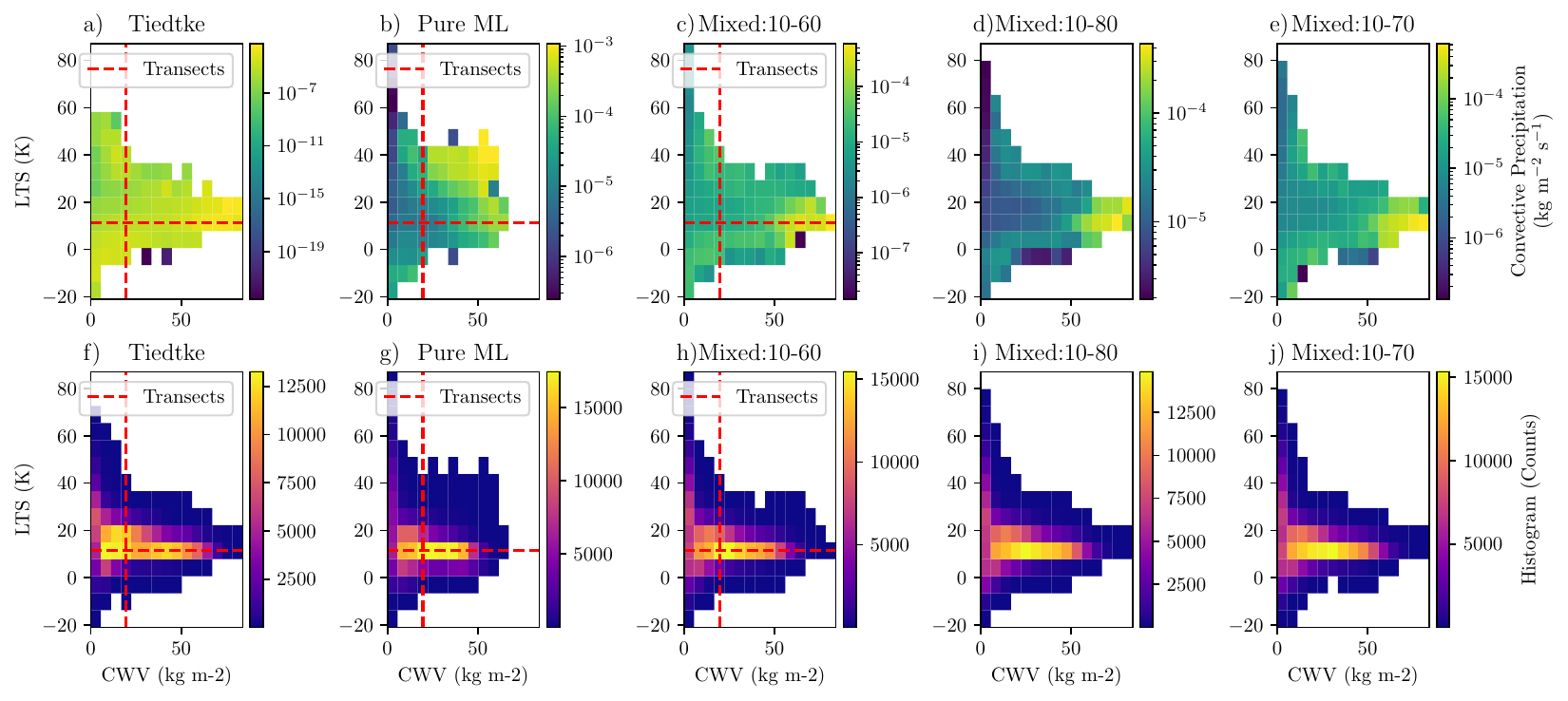}
    \caption{2D histogram of LTS and CWV for 5 different coupled schemes in the top row (a-e). Additionally, the conditionally averaged convective precipitation for each bin above as a function of LTS and CWV is displayed in the lower row (f-j).}
    \label{fig:precipitation_prw_lts_2d}
\end{figure}

\begin{figure}[tbh]
    \centering
    \includegraphics[width=1\linewidth]{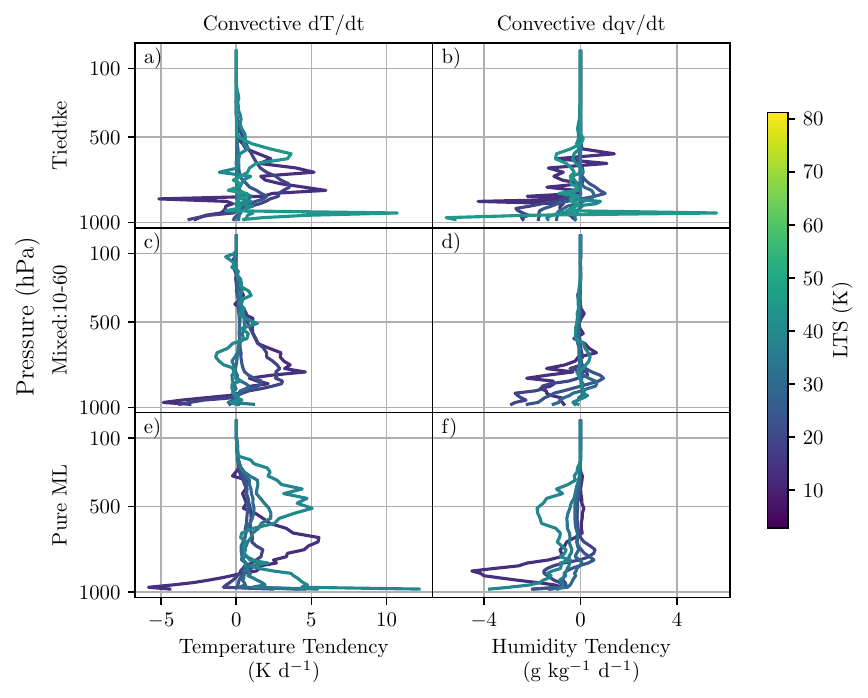}
    \caption{Conditional averages of convective heating rates (first column) and moistening rates (second column) as a function of height. The conditioning is based on LTS while we keep the value for the CWV fixed to $CWV=\qty{19.6}{\kilogram\per\meter^2}$. Each row corresponds to a different coupled scheme: (a,b) for Tiedtke, (c,d) for Mixed:10-60, and (e,f) for the pure ML scheme. Conditional averaged curves are only computed for LTS conditions having at least ten samples.}
       \label{fig:1d_average-profiles_fixedcwv}
   \end{figure}

\clearpage

\begin{figure}[tbh]
    \centering
    \includegraphics[width=0.96\linewidth]{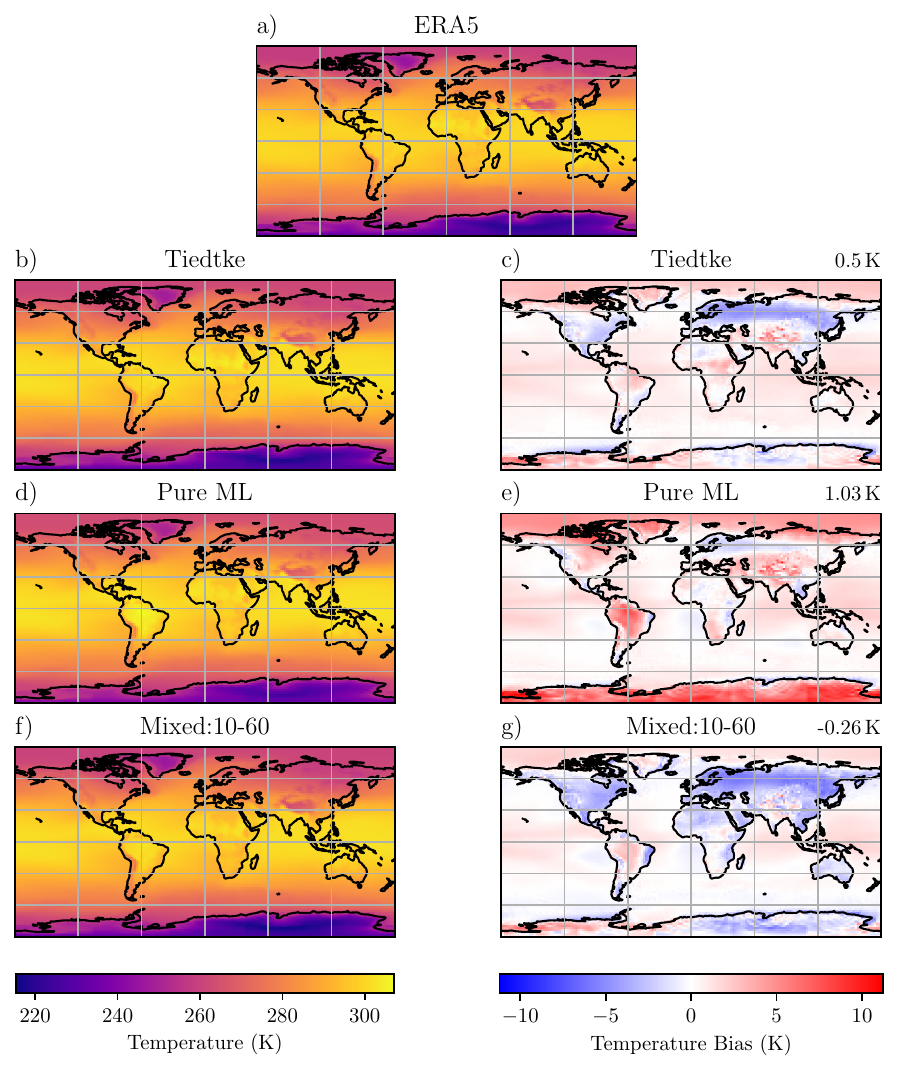}
    \caption{Spatial distribution of 20-year averaged near surface temperature $T_{2m}$ for different convection schemes in the left column and the bias with respect to ERA5 in the right column. The first row (a) shows near surface temperature for the ERA5 data, the Tiedtke scheme in the second row (b-c), the pure ML scheme in the third row (d-e), and the Mixed:10-60 scheme in the last row (f-g). In the upper right of each bias plot, the area-weighted mean bias is displayed.}
    \label{fig:20y_tas_vs_era5}
\end{figure}

\begin{table}[htb]
    \centering
    \begin{tabular}{l l l l}
        \toprule
        Metric - Variable & Tiedtke & Pure ML & Mixed:10-60\\
        \midrule
        Mean Bias - $T_{2m}$ (\unit{\kelvin}) & 0.50 & 1.03 & \textbf{-0.26}\\
        RMSE - $T_{2m}$ (\unit{\kelvin}) & \textbf{1.37} & 1.88 & 1.54\\
        MAE - $T_{2m}$ (\unit{\kelvin}) & 1.04 & 1.22 & \textbf{1.03}\\
        Mean Bias - Precipitation (\unit{\milli\meter\per\day}) & -0.21 & -0.34 & \textbf{0.14}\\
        RMSE - Precipitation (\unit{\milli\meter\per\day}) & \textbf{1.35} & 1.36 & 1.48\\
        MAE - Precipitation (\unit{\milli\meter\per\day}) & \textbf{0.86} & 0.87 & 0.96\\
        \bottomrule
    \end{tabular}
    \caption{The area weighted mean bias, RMSE, and MAE for near-surface Temperature and Precipitation corresponding to \Cref{fig:20y_tas_vs_era5,fig:20y_precip_vs_gpcp}.}
    \label{tab:spatial_mean_bias_20_year}
\end{table}

\clearpage

\section*{Open Research}
The code will be published under \url{https://github.com/EyringMLClimateGroup/heuer25_ml_convection_climsim} and preserved \cite{helgehr_2025_17234569}.
All training data is openly accessible under \citeA{ai_climsim_high-res_2023}.
The software code for the ICON model is available from \url{https://www.icon-model.org/}.

\acknowledgments
Helge Heuer, Julien Savre, Manuel Schlund, and Veronika Eyring received funding for this study from the European Research Council (ERC) Synergy Grant ``Understanding and Modelling the Earth System with Machine Learning (USMILE)'' under the Horizon 2020 research and innovation programme (Grant agreement No. 855187) and from the Horizon Europe project ``Artificial Intelligence for enhanced representation of processes and extremes in Earth System Models (AI4PEX)” (Grant agreement No. 101137682).
Tom Beucler received support from AIPEX, funded by the Swiss State Secretariat for Education, Research and Innovation (SERI, Grant No. 23.00546).
The contribution by Mierk Schwabe was made possible by the DLR Quantum Computing Initiative and the Federal Ministry for Economic Affairs and Climate Action; \url{qci.dlr.de/projects/klim-qml}.
This work used resources of the Deutsches Klimarechenzentrum (DKRZ) granted by its Scientific Steering Committee (WLA) under project ID bd1179.
The authors gratefully acknowledge the Earth System Modelling Project (ESM) for funding this work by providing computing time on the ESM partition of the supercomputer JUWELS \cite{JUWELS} at the Jülich Supercomputing Centre (JSC).
We thank the authors of \citeA{yu_climsim_2023} for creating and providing the global E3SM-MMF simulations used in this study.
Furthermore, we thank the hosts of the Kaggle competition \citeA{lin_leap_2024} and all their participants for providing highly competitive ML baselines for the parameterization problem and bringing the ML and atmospheric science communities closer together.
We also thank the three anonymous reviewers for their thorough evaluation and constructive comments.

\bibliography{references}

@article{brenowitz_interpreting_2020,
	title = {Interpreting and stabilizing machine-learning parametrizations of convection},
	volume = {77},
	issn = {0022-4928},
	number = {12},
	journal = {Journal of the atmospheric sciences},
	author = {Brenowitz, Noah D. and Beucler, Tom and Pritchard, Michael and Bretherton, Christopher S.},
	year = {2020},
	pages = {4357--4375},
	}

@article{tiedtke_comprehensive_1989,
	title = {A comprehensive mass flux scheme for cumulus parameterization in large-scale models},
	volume = {117},
	issn = {1520-0493},
	number = {8},
	journal = {Monthly weather review},
	author = {Tiedtke, MICHAEL},
	year = {1989},
	pages = {1779--1800},
	}

@article{nordeng_extended_1994,
	title = {Extended versions of the convective parametrization scheme at {ECMWF} and their impact on the mean and transient activity of the model in the tropics},
	volume = {206},
	journal = {Research Department Technical Memorandum},
	author = {Nordeng, Thor Erik},
	year = {1994},
	pages = {1--41},
	}

@article{rasp_deep_2018,
	title = {Deep learning to represent subgrid processes in climate models},
	volume = {115},
	issn = {0027-8424},
	number = {39},
	journal = {Proceedings of the National Academy of Sciences},
	author = {Rasp, Stephan and Pritchard, Michael S. and Gentine, Pierre},
	year = {2018},
	pages = {9684--9689},
	}

@article{yuval_stable_2020,
	title = {Stable machine-learning parameterization of subgrid processes for climate modeling at a range of resolutions},
	volume = {11},
	issn = {2041-1723},
	number = {1},
	journal = {Nature communications},
	author = {Yuval, Janni and O’Gorman, Paul A.},
	year = {2020},
	pages = {1--10},
	}

@article{gentine_could_2018,
	title = {Could {Machine} {Learning} {Break} the {Convection} {Parameterization} {Deadlock}?},
	volume = {45},
	issn = {0094-8276},
	doi = {https://doi.org/10.1029/2018GL078202},
	number = {11},
	journal = {Geophysical Research Letters},
	author = {Gentine, P. and Pritchard, M. and Rasp, S. and Reinaudi, G. and Yacalis, G.},
	year = {2018},
	pages = {5742--5751},
	}

@article{christopoulos_assessing_2021,
	title = {Assessing {Biases} and {Climate} {Implications} of the {Diurnal} {Precipitation} {Cycle} in {Climate} {Models}},
	volume = {48},
	issn = {0094-8276},
	doi = {https://doi.org/10.1029/2021GL093017},
	number = {13},
	journal = {Geophysical Research Letters},
	author = {Christopoulos, Costa and Schneider, Tapio},
	year = {2021},
	pages = {e2021GL093017},
	}

@article{randall_breaking_2003,
	title = {Breaking the cloud parameterization deadlock},
	volume = {84},
	issn = {0003-0007},
	number = {11},
	journal = {Bulletin of the American Meteorological Society},
	author = {Randall, David and Khairoutdinov, Marat and Arakawa, Akio and Grabowski, Wojciech},
	year = {2003},
	pages = {1547--1564},
	}

@article{colin_identifying_2019,
	title = {Identifying the sources of convective memory in cloud-resolving simulations},
	volume = {76},
	issn = {0022-4928},
	number = {3},
	journal = {Journal of the atmospheric sciences},
	author = {Colin, Maxime and Sherwood, Steven and Geoffroy, Olivier and Bony, Sandrine and Fuchs, David},
	year = {2019},
	pages = {947--962},
	}

@article{arakawa_interaction_1974,
	title = {Interaction of a cumulus cloud ensemble with the large-scale environment, {Part} {I}},
	volume = {31},
	issn = {0022-4928},
	number = {3},
	journal = {Journal of the atmospheric sciences},
	author = {Arakawa, Akio and Schubert, Wayne Howard},
	year = {1974},
	pages = {674--701},
	}

@article{giorgetta_icon-model_2022,
	title = {The {ICON}-{A} model for direct {QBO} simulations on {GPUs} (version icon-cscs: baf28a514)},
	journal = {EGUsphere},
	author = {Giorgetta, Marco A. and Sawyer, William and Lapillonne, Xavier and Adamidis, Panagiotis and Alexeev, Dmitry and Clément, Valentin and Dietlicher, Remo and Engels, Jan Frederik and Esch, Monika and Franke, Henning},
	year = {2022},
	pages = {1--46},
	}

@article{stevens_dyamond_2019,
	title = {{DYAMOND}: the {DYnamics} of the {Atmospheric} general circulation {Modeled} {On} {Non}-hydrostatic {Domains}},
	volume = {6},
	issn = {2197-4284},
	number = {1},
	journal = {Progress in Earth and Planetary Science},
	author = {Stevens, Bjorn and Satoh, Masaki and Auger, Ludovic and Biercamp, Joachim and Bretherton, Christopher S. and Chen, Xi and Düben, Peter and Judt, Falko and Khairoutdinov, Marat and Klocke, Daniel},
	year = {2019},
	pages = {1--17},
	}

@article{stephens_dreary_2010,
	title = {Dreary state of precipitation in global models},
	volume = {115},
	issn = {0148-0227},
	doi = {https://doi.org/10.1029/2010JD014532},
	number = {D24},
	journal = {Journal of Geophysical Research: Atmospheres},
	author = {Stephens, Graeme L. and L'Ecuyer, Tristan and Forbes, Richard and Gettelmen, Andrew and Golaz, Jean-Christophe and Bodas-Salcedo, Alejandro and Suzuki, Kentaroh and Gabriel, Philip and Haynes, John},
	year = {2010},
	}

@article{giorgetta_icon-_2018,
	title = {{ICON}-{A}, the {Atmosphere} {Component} of the {ICON} {Earth} {System} {Model}: {I}. {Model} {Description}},
	volume = {10},
	issn = {1942-2466},
	doi = {https://doi.org/10.1029/2017MS001242},
	number = {7},
	journal = {Journal of Advances in Modeling Earth Systems},
	author = {Giorgetta, M. A. and Brokopf, R. and Crueger, T. and Esch, M. and Fiedler, S. and Helmert, J. and Hohenegger, C. and Kornblueh, L. and Köhler, M. and Manzini, E. and Mauritsen, T. and Nam, C. and Raddatz, T. and Rast, S. and Reinert, D. and Sakradzija, M. and Schmidt, H. and Schneck, R. and Schnur, R. and Silvers, L. and Wan, H. and Zängl, G. and Stevens, B.},
	year = {2018},
	pages = {1613--1637},
	}

@article{brenowitz_prognostic_2018,
	title = {Prognostic validation of a neural network unified physics parameterization},
	volume = {45},
	issn = {0094-8276},
	number = {12},
	journal = {Geophysical Research Letters},
	author = {Brenowitz, Noah D. and Bretherton, Christopher S.},
	year = {2018},
	pages = {6289--6298},
	}

@article{watt-meyer_neural_2024,
	title = {Neural {Network} {Parameterization} of {Subgrid}-{Scale} {Physics} {From} a {Realistic} {Geography} {Global} {Storm}-{Resolving} {Simulation}},
	volume = {16},
	issn = {1942-2466},
	doi = {https://doi.org/10.1029/2023MS003668},
	number = {2},
	journal = {Journal of Advances in Modeling Earth Systems},
	author = {Watt-Meyer, Oliver and Brenowitz, Noah D. and Clark, Spencer K. and Henn, Brian and Kwa, Anna and McGibbon, Jeremy and Perkins, W. Andre and Harris, Lucas and Bretherton, Christopher S.},
	year = {2024},
	pages = {e2023MS003668},
	}

@article{brenowitz_spatially_2019,
	title = {Spatially {Extended} {Tests} of a {Neural} {Network} {Parametrization} {Trained} by {Coarse}-{Graining}},
	volume = {11},
	issn = {1942-2466},
	doi = {https://doi.org/10.1029/2019MS001711},
	number = {8},
	journal = {Journal of Advances in Modeling Earth Systems},
	author = {Brenowitz, Noah D. and Bretherton, Christopher S.},
	year = {2019},
	pages = {2728--2744},
	}

@article{yuval_neural-network_2023,
	title = {Neural-{Network} {Parameterization} of {Subgrid} {Momentum} {Transport} in the {Atmosphere}},
	volume = {15},
	issn = {1942-2466},
	doi = {https://doi.org/10.1029/2023MS003606},
	number = {4},
	journal = {Journal of Advances in Modeling Earth Systems},
	author = {Yuval, Janni and O’Gorman, Paul A.},
	year = {2023},
	pages = {e2023MS003606},
	}

@article{hu2025stable,
  title={Stable machine-learning parameterization of subgrid processes in a comprehensive atmospheric model learned from embedded convection-permitting simulations},
  author={Hu, Zeyuan and Subramaniam, Akshay and Kuang, Zhiming and Lin, Jerry and Yu, Sungduk and Hannah, Walter M and Brenowitz, Noah D and Romero, Josh and Pritchard, Michael S},
  journal={Journal of Advances in Modeling Earth Systems},
  volume={17},
  number={7},
  pages={e2024MS004618},
  year={2025},
  publisher={Wiley Online Library}
}

@misc{yu_climsim-online_2024,
	title = {{ClimSim}-{Online} v6: {A} {Large} {Multi}-scale {Dataset} and {Framework} for {Hybrid} {ML}-physics {Climate} {Emulation}},
	testurl = {https://arxiv.org/abs/2306.08754},
	author = {Yu, Sungduk and Hu, Zeyuan and Subramaniam, Akshay and Hannah, Walter and Peng, Liran and Lin, Jerry and Bhouri, Mohamed Aziz and Gupta, Ritwik and Lütjens, Björn and Will, Justus C. and Behrens, Gunnar and Busecke, Julius J. M. and Loose, Nora and Stern, Charles I. and Beucler, Tom and Harrop, Bryce and Heuer, Helge and Hillman, Benjamin R. and Jenney, Andrea and Liu, Nana and White, Alistair and Zheng, Tian and Kuang, Zhiming and Ahmed, Fiaz and Barnes, Elizabeth and Brenowitz, Noah D. and Bretherton, Christopher and Eyring, Veronika and Ferretti, Savannah and Lutsko, Nicholas and Gentine, Pierre and Mandt, Stephan and Neelin, J. David and Yu, Rose and Zanna, Laure and Urban, Nathan and Yuval, Janni and Abernathey, Ryan and Baldi, Pierre and Chuang, Wayne and Huang, Yu and Iglesias-Suarez, Fernando and Jantre, Sanket and Ma, Po-Lun and Shamekh, Sara and Zhang, Guang and Pritchard, Michael},
	  year = {2024},
	testnote = {\_eprint: 2306.08754},
	}

@article{JMLR:v26:24-1014,
  author  = {Sungduk Yu and Zeyuan Hu and Akshay Subramaniam and Walter Hannah and Liran Peng and Jerry Lin and Mohamed Aziz Bhouri and Ritwik Gupta and Bj{{\"o}}rn L{{\"u}}tjens and Justus C. Will and Gunnar Behrens and Julius J. M. Busecke and Nora Loose and Charles I Stern and Tom Beucler and Bryce Harrop and Helge Heuer and Benjamin R Hillman and Andrea Jenney and Nana Liu and Alistair White and Tian Zheng and Zhiming Kuang and Fiaz Ahmed and Elizabeth Barnes and Noah D. Brenowitz and Christopher Bretherton and Veronika Eyring and Savannah Ferretti and Nicholas Lutsko and Pierre Gentine and Stephan Mandt and J. David Neelin and Rose Yu and Laure Zanna and Nathan M. Urban and Janni Yuval and Ryan Abernathey and Pierre Baldi and Wayne Chuang and Yu Huang and Fernando Iglesias-Suarez and Sanket Jantre and Po-Lun Ma and Sara Shamekh and Guang Zhang and Michael Pritchard},
  title   = {ClimSim-Online: A Large Multi-Scale Dataset and Framework for Hybrid Physics-ML Climate Emulation},
  journal = {Journal of Machine Learning Research},
  year    = {2025},
  volume  = {26},
  number  = {142},
  pages   = {1--85},
  testurl     = {http://jmlr.org/papers/v26/24-1014.html}
}

@article{sanford_improving_2023,
	title = {Improving the {Reliability} of {ML}-{Corrected} {Climate} {Models} {With} {Novelty} {Detection}},
	volume = {15},
	testurl = {https://agupubs.onlinelibrary.wiley.com/doi/abs/10.1029/2023MS003809},
	doi = {https://doi.org/10.1029/2023MS003809},
	number = {11},
	journal = {Journal of Advances in Modeling Earth Systems},
	author = {Sanford, Clayton and Kwa, Anna and Watt-Meyer, Oliver and Clark, Spencer K. and Brenowitz, Noah and McGibbon, Jeremy and Bretherton, Christopher},
	year = {2023},
	testnote = {\_eprint: https://agupubs.onlinelibrary.wiley.com/doi/pdf/10.1029/2023MS003809},
	pages = {e2023MS003809},
	}

@article{lin2025navigating,
  title={Navigating the Noise: Bringing Clarity to ML Parameterization Design With O O (100) Ensembles},
  author={Lin, Jerry and Yu, Sungduk and Peng, Liran and Beucler, Tom and Wong-Toi, Eliot and Hu, Zeyuan and Gentine, Pierre and Geleta, Margarita and Pritchard, Mike},
  journal={Journal of Advances in Modeling Earth Systems},
  volume={17},
  number={4},
  pages={e2024MS004551},
  year={2025},
  publisher={Wiley Online Library}
}

@article{heuer_interpretable_2024,
	title = {Interpretable {Multiscale} {Machine} {Learning}-{Based} {Parameterizations} of {Convection} for {ICON}},
	volume = {16},
	testurl = {https://agupubs.onlinelibrary.wiley.com/doi/abs/10.1029/2024MS004398},
	doi = {https://doi.org/10.1029/2024MS004398},
	number = {8},
	journal = {Journal of Advances in Modeling Earth Systems},
	author = {Heuer, Helge and Schwabe, Mierk and Gentine, Pierre and Giorgetta, Marco A. and Eyring, Veronika},
	year = {2024},
	testnote = {\_eprint: https://agupubs.onlinelibrary.wiley.com/doi/pdf/10.1029/2024MS004398},
	pages = {e2024MS004398},
	}

@article{pincus_balancing_2019,
	title = {Balancing {Accuracy}, {Efficiency}, and {Flexibility} in {Radiation} {Calculations} for {Dynamical} {Models}},
	volume = {11},
	testurl = {https://agupubs.onlinelibrary.wiley.com/doi/abs/10.1029/2019MS001621},
	doi = {https://doi.org/10.1029/2019MS001621},
	number = {10},
	journal = {Journal of Advances in Modeling Earth Systems},
	author = {Pincus, Robert and Mlawer, Eli J. and Delamere, Jennifer S.},
	year = {2019},
	testnote = {\_eprint: https://agupubs.onlinelibrary.wiley.com/doi/pdf/10.1029/2019MS001621},
	pages = {3074--3089},
	}

@misc{pincus_rterrtmgp_2023,
	title = {{RTE}+{RRTMGP}},
	url = {https://github.com/earth-system-radiaton/rte-rrtmgp},
	author = {Pincus, Robert and Iacono, Michael J. and Alexeev, Dmitry and Adamidis, Panos and Hillman, Benjamin R. and Norman, Matthew and Pfister, Erik and Polonsky, Igor N. and Romero, Nicols A. and Kosukhin, Sergey S. and Wehe, Andre},
	month = nov,
	year = {2023},
}

@misc{ai_climsim_high-res_2023,
	title = {{ClimSim}\_high-res ({Revision} d251368)},
	url = {https://huggingface.co/datasets/LEAP/ClimSim_high-res},
	publisher = {Hugging Face},
	author = {LEAP},
	year = {2023},
	doi = {10.57967/hf/0739},
}

@misc{e3sm_project_energy_2018,
	title = {Energy {Exascale} {Earth} {System} {Model} ({E3SM})},
	testurl = {https://dx.doi.org/10.11578/E3SM/dc.20180418.36},
	author = {{E3SM Project}},
	month = apr,
	year = {2018},
	doi = {10.11578/E3SM/dc.20180418.36},
	testnote = {Published: [Computer Software] https://dx.doi.org/10.11578/E3SM/dc.20180418.36},
}

@article{liaw_tune_2018,
	title = {Tune: {A} {Research} {Platform} for {Distributed} {Model} {Selection} and {Training}},
	journal = {arXiv preprint arXiv:1807.05118},
	author = {Liaw, Richard and Liang, Eric and Nishihara, Robert and Moritz, Philipp and Gonzalez, Joseph E and Stoica, Ion},
	year = {2018},
}

@misc{andela_esmvaltool_2025,
	title = {{ESMValTool}},
	url = {https://github.com/ESMValGroup/ESMValTool/},
	author = {Andela, Bouwe and Broetz, Bjoern and de Mora, Lee and Drost, Niels and Eyring, Veronika and Koldunov, Nikolay and Lauer, Axel and Mueller, Benjamin and Predoi, Valeriu and Righi, Mattia and Schlund, Manuel and Vegas-Regidor, Javier and Zimmermann, Klaus and Adeniyi, Kemisola and Arnone, Enrico and Bellprat, Omar and Berg, Peter and Billows, Chris and Bock, Lisa and Bodas-Salcedo, Alejandro and Caron, Louis-Philippe and Carvalhais, Nuno and Cionni, Irene and Cortesi, Nicola and Corti, Susanna and Crezee, Bas and Davin, Edouard Leopold and Davini, Paolo and Deser, Clara and Diblen, Faruk and Docquier, David and Dreyer, Laura and Ehbrecht, Carsten and Earnshaw, Paul and Geddes, Theo and Gier, Bettina and Gillett, Ed and Gonzalez-Reviriego, Nube and Goodman, Paul and Hagemann, Stefan and Hardacre, Catherine and von Hardenberg, Jost and Hassler, Birgit and Heuer, Helge and Hogan, Emma and Hunter, Alasdair and Kadow, Christopher and Kindermann, Stephan and Koirala, Sujan and Kuehbacher, Birgit and Lledó, Llorenç and Lejeune, Quentin and Lembo, Valerio and Little, Bill and Loosveldt-Tomas, Saskia and Lorenz, Ruth and Lovato, Tomas and Lucarini, Valerio and Malinina, Elizaveta and Massonnet, François and Mohr, Christian Wilhelm and Amarjiit, Pandde and Pérez-Zanón, Núria and Phillips, Adam and Proft, Max and Russell, Joellen and Sandstad, Marit and Sellar, Alistair and Senftleben, Daniel and Serva, Federico and Sillmann, Jana and Stacke, Tobias and Swaminathan, Ranjini and Tomkins, Katherine and Torralba, Verónica and Weigel, Katja and Sarauer, Ellen and Roberts, Charles and Kalverla, Peter and Alidoost, Sarah and Verhoeven, Stefan and Vreede, Barbara and Smeets, Stef and Soares Siqueira, Abel and Kazeroni, Rémi and Potter, Jerry and Winterstein, Franziska and Beucher, Romain and Kraft, Jeremy and Ruhe, Lukas and Bonnet, Pauline and Munday, Gregory and Chun, Felicity},
	month = mar,
	year = {2025},
	doi = {10.5281/zenodo.3401363},
}

@misc{lin_leap_2024,
	title = {{LEAP} - {Atmospheric} {Physics} using {AI} ({ClimSim})},
	url = {https://kaggle.com/competitions/leap-atmospheric-physics-ai-climsim},
	author = {Lin, Jerry and Hu, Zeyuan and Yu, Sungduk and Pritchard, Mike and Gupta, Ritwik and Zheng, Tian and Hannah, Walter and Mansfield, Laura and Qu, Yongquan and Geleta, Margarita and Lopez, Molly and Rudolph, Maja and Chow, Ashley and Reade, Walter},
	year = {2024},
}

@article{beucler_distilling_2025,
	title = {Distilling {Machine} {Learning}’s {Added} {Value}: {Pareto} {Fronts} in {Atmospheric} {Applications}},
	volume = {4},
	testurl = {https://journals.ametsoc.org/view/journals/aies/4/2/AIES-D-24-0078.1.xml},
	doi = {10.1175/AIES-D-24-0078.1},
	number = {2},
	journal = {Artificial Intelligence for the Earth Systems},
	author = {Beucler, Tom and Grundner, Arthur and Shamekh, Sara and Ukkonen, Peter and Chantry, Matthew and Lagerquist, Ryan},
	year = {2025},
	testnote = {Place: Boston MA, USA
Publisher: American Meteorological Society},
	pages = {e240078},
}

@article{shen_engression_2024,
	title = {Engression: extrapolation through the lens of distributional regression},
	issn = {1369-7412},
	testurl = {https://doi.org/10.1093/jrsssb/qkae108},
	doi = {10.1093/jrsssb/qkae108},
	journal = {Journal of the Royal Statistical Society Series B: Statistical Methodology},
	author = {Shen, Xinwei and Meinshausen, Nicolai},
	month = nov,
	year = {2024},
	testnote = {\_eprint: https://academic.oup.com/jrsssb/advance-article-pdf/doi/10.1093/jrsssb/qkae108/60827977/qkae108.pdf},
	pages = {qkae108},
}

@article{adler_global_2018,
	title = {The {Global} {Precipitation} {Climatology} {Project} ({GPCP}) {Monthly} {Analysis} ({New} {Version} 2.3) and a {Review} of 2017 {Global} {Precipitation}},
	volume = {9},
	issn = {2073-4433},
	testurl = {https://www.mdpi.com/2073-4433/9/4/138},
	doi = {10.3390/atmos9040138},
	number = {4},
	journal = {Atmosphere},
	author = {Adler, Robert F. and Sapiano, Mathew R. P. and Huffman, George J. and Wang, Jian-Jian and Gu, Guojun and Bolvin, David and Chiu, Long and Schneider, Udo and Becker, Andreas and Nelkin, Eric and Xie, Pingping and Ferraro, Ralph and Shin, Dong-Bin},
	year = {2018},
}

@article{gelaro_modern-era_2017,
	title = {The {Modern}-{Era} {Retrospective} {Analysis} for {Research} and {Applications}, {Version} 2 ({MERRA}-2)},
	volume = {30},
	testurl = {https://journals.ametsoc.org/view/journals/clim/30/14/jcli-d-16-0758.1.xml},
	doi = {10.1175/JCLI-D-16-0758.1},
	number = {14},
	journal = {Journal of Climate},
	author = {Gelaro, Ronald and McCarty, Will and Suárez, Max J. and Todling, Ricardo and Molod, Andrea and Takacs, Lawrence and Randles, Cynthia A. and Darmenov, Anton and Bosilovich, Michael G. and Reichle, Rolf and Wargan, Krzysztof and Coy, Lawrence and Cullather, Richard and Draper, Clara and Akella, Santha and Buchard, Virginie and Conaty, Austin and Silva, Arlindo M. da and Gu, Wei and Kim, Gi-Kong and Koster, Randal and Lucchesi, Robert and Merkova, Dagmar and Nielsen, Jon Eric and Partyka, Gary and Pawson, Steven and Putman, William and Rienecker, Michele and Schubert, Siegfried D. and Sienkiewicz, Meta and Zhao, Bin},
	year = {2017},
	testnote = {Place: Boston MA, USA
Publisher: American Meteorological Society},
	pages = {5419 -- 5454},
}

@article{hersbach_era5_2020,
	title = {The {ERA5} global reanalysis},
	volume = {146},
	testurl = {https://rmets.onlinelibrary.wiley.com/doi/abs/10.1002/qj.3803},
	doi = {https://doi.org/10.1002/qj.3803},
	number = {730},
	journal = {Quarterly Journal of the Royal Meteorological Society},
	author = {Hersbach, Hans and Bell, Bill and Berrisford, Paul and Hirahara, Shoji and Horányi, András and Muñoz-Sabater, Joaquín and Nicolas, Julien and Peubey, Carole and Radu, Raluca and Schepers, Dinand and Simmons, Adrian and Soci, Cornel and Abdalla, Saleh and Abellan, Xavier and Balsamo, Gianpaolo and Bechtold, Peter and Biavati, Gionata and Bidlot, Jean and Bonavita, Massimo and De Chiara, Giovanna and Dahlgren, Per and Dee, Dick and Diamantakis, Michail and Dragani, Rossana and Flemming, Johannes and Forbes, Richard and Fuentes, Manuel and Geer, Alan and Haimberger, Leo and Healy, Sean and Hogan, Robin J. and Hólm, Elías and Janisková, Marta and Keeley, Sarah and Laloyaux, Patrick and Lopez, Philippe and Lupu, Cristina and Radnoti, Gabor and de Rosnay, Patricia and Rozum, Iryna and Vamborg, Freja and Villaume, Sebastien and Thépaut, Jean-Noël},
	year = {2020},
	testnote = {\_eprint: https://rmets.onlinelibrary.wiley.com/doi/pdf/10.1002/qj.3803},
	pages = {1999--2049},
}

@misc{grundner_reduced_2025,
	title = {Reduced {Cloud} {Cover} {Errors} in a {Hybrid} {AI}-{Climate} {Model} {Through} {Equation} {Discovery} {And} {Automatic} {Tuning}},
	testurl = {https://arxiv.org/abs/2505.04358},
	author = {Grundner, Arthur and Beucler, Tom and Savre, Julien and Lauer, Axel and Schlund, Manuel and Eyring, Veronika},
	year = {2025},
	testnote = {\_eprint: 2505.04358},
	}

@inproceedings{ansel_pytorch_2024,
	title = {{PyTorch} 2: {Faster} {Machine} {Learning} {Through} {Dynamic} {Python} {Bytecode} {Transformation} and {Graph} {Compilation}},
	testurl = {https://docs.pytorch.org/assets/pytorch2-2.pdf},
	doi = {10.1145/3620665.3640366},
	booktitle = {29th {ACM} {International} {Conference} on {Architectural} {Support} for {Programming} {Languages} and {Operating} {Systems}, {Volume} 2 ({ASPLOS} '24)},
	publisher = {ACM},
	author = {Ansel, Jason and Yang, Edward and He, Horace and Gimelshein, Natalia and Jain, Animesh and Voznesensky, Michael and Bao, Bin and Bell, Peter and Berard, David and Burovski, Evgeni and Chauhan, Geeta and Chourdia, Anjali and Constable, Will and Desmaison, Alban and DeVito, Zachary and Ellison, Elias and Feng, Will and Gong, Jiong and Gschwind, Michael and Hirsh, Brian and Huang, Sherlock and Kalambarkar, Kshiteej and Kirsch, Laurent and Lazos, Michael and Lezcano, Mario and Liang, Yanbo and Liang, Jason and Lu, Yinghai and Luk, CK and Maher, Bert and Pan, Yunjie and Puhrsch, Christian and Reso, Matthias and Saroufim, Mark and Siraichi, Marcos Yukio and Suk, Helen and Suo, Michael and Tillet, Phil and Wang, Eikan and Wang, Xiaodong and Wen, William and Zhang, Shunting and Zhao, Xu and Zhou, Keren and Zou, Richard and Mathews, Ajit and Chanan, Gregory and Wu, Peng and Chintala, Soumith},
	month = apr,
	year = {2024},
}

@misc{loshchilov_decoupled_2019,
	title = {Decoupled {Weight} {Decay} {Regularization}},
	testurl = {https://arxiv.org/abs/1711.05101},
	author = {Loshchilov, Ilya and Hutter, Frank},
	year = {2019},
	testnote = {\_eprint: 1711.05101},
}

@article{judt_insights_2018,
	title = {Insights into {Atmospheric} {Predictability} through {Global} {Convection}-{Permitting} {Model} {Simulations}},
	volume = {75},
	testurl = {https://journals.ametsoc.org/view/journals/atsc/75/5/jas-d-17-0343.1.xml},
	doi = {10.1175/JAS-D-17-0343.1},
	number = {5},
	journal = {Journal of the Atmospheric Sciences},
	author = {Judt, Falko},
	year = {2018},
	testnote = {Place: Boston MA, USA
Publisher: American Meteorological Society},
	pages = {1477 -- 1497},
}

@article{grundner_deep_2022,
	title = {Deep {Learning} {Based} {Cloud} {Cover} {Parameterization} for {ICON}},
	volume = {14},
	testurl = {https://agupubs.onlinelibrary.wiley.com/doi/abs/10.1029/2021MS002959},
	doi = {https://doi.org/10.1029/2021MS002959},
	number = {12},
	journal = {Journal of Advances in Modeling Earth Systems},
	author = {Grundner, Arthur and Beucler, Tom and Gentine, Pierre and Iglesias-Suarez, Fernando and Giorgetta, Marco A. and Eyring, Veronika},
	year = {2022},
	testnote = {\_eprint: https://agupubs.onlinelibrary.wiley.com/doi/pdf/10.1029/2021MS002959},
	pages = {e2021MS002959},
}

@article{grundner_data-driven_2024,
	title = {Data-{Driven} {Equation} {Discovery} of a {Cloud} {Cover} {Parameterization}},
	volume = {16},
	testurl = {https://agupubs.onlinelibrary.wiley.com/doi/abs/10.1029/2023MS003763},
	doi = {https://doi.org/10.1029/2023MS003763},
	number = {3},
	journal = {Journal of Advances in Modeling Earth Systems},
	author = {Grundner, Arthur and Beucler, Tom and Gentine, Pierre and Eyring, Veronika},
	year = {2024},
	testnote = {\_eprint: https://agupubs.onlinelibrary.wiley.com/doi/pdf/10.1029/2023MS003763},
	pages = {e2023MS003763},
}

@article{sukovich_extreme_2014,
	title = {Extreme {Quantitative} {Precipitation} {Forecast} {Performance} at the {Weather} {Prediction} {Center} from 2001 to 2011},
	volume = {29},
	testurl = {https://journals.ametsoc.org/view/journals/wefo/29/4/waf-d-13-00061_1.xml},
	doi = {10.1175/WAF-D-13-00061.1},
	number = {4},
	journal = {Weather and Forecasting},
	author = {Sukovich, Ellen M. and Ralph, F. Martin and Barthold, Faye E. and Reynolds, David W. and Novak, David R.},
	year = {2014},
	testnote = {Place: Boston MA, USA
Publisher: American Meteorological Society},
	pages = {894 -- 911},
}

@article{jones_global_2004,
	title = {Global {Occurrences} of {Extreme} {Precipitation} and the {Madden}–{Julian} {Oscillation}: {Observations} and {Predictability}},
	volume = {17},
	testurl = {https://journals.ametsoc.org/view/journals/clim/17/23/3238.1.xml},
	doi = {10.1175/3238.1},
	number = {23},
	journal = {Journal of Climate},
	author = {Jones, Charles and Waliser, Duane E. and Lau, K. M. and Stern, W.},
	year = {2004},
	testnote = {Place: Boston MA, USA
Publisher: American Meteorological Society},
	pages = {4575 -- 4589},
}

@article{hannah_checkerboard_2022,
	title = {Checkerboard patterns in {E3SMv2} and {E3SM}-{MMFv2}},
	volume = {15},
	testurl = {https://gmd.copernicus.org/articles/15/6243/2022/},
	doi = {10.5194/gmd-15-6243-2022},
	number = {15},
	journal = {Geoscientific Model Development},
	author = {Hannah, W. and Pressel, K. and Ovchinnikov, M. and Elsaesser, G.},
	year = {2022},
	pages = {6243--6257},
}

@inproceedings{yu_climsim_2023,
	title = {{ClimSim}: {A} large multi-scale dataset for hybrid physics-{ML} climate emulation},
	volume = {36},
	testurl = {https://proceedings.neurips.cc/paper_files/paper/2023/file/45fbcc01349292f5e059a0b8b02c8c3f-Paper-Datasets_and_Benchmarks.pdf},
	booktitle = {Advances in {Neural} {Information} {Processing} {Systems}},
	publisher = {Curran Associates, Inc.},
	author = {Yu, Sungduk and Hannah, Walter and Peng, Liran and Lin, Jerry and Bhouri, Mohamed Aziz and Gupta, Ritwik and Lütjens, Björn and Will, Justus C. and Behrens, Gunnar and Busecke, Julius and Loose, Nora and Stern, Charles and Beucler, Tom and Harrop, Bryce and Hillman, Benjamin and Jenney, Andrea and Ferretti, Savannah L. and Liu, Nana and Anandkumar, Animashree and Brenowitz, Noah and Eyring, Veronika and Geneva, Nicholas and Gentine, Pierre and Mandt, Stephan and Pathak, Jaideep and Subramaniam, Akshay and Vondrick, Carl and Yu, Rose and Zanna, Laure and Zheng, Tian and Abernathey, Ryan and Ahmed, Fiaz and Bader, David and Baldi, Pierre and Barnes, Elizabeth and Bretherton, Christopher and Caldwell, Peter and Chuang, Wayne and Han, Yilun and HUANG, YU and Iglesias-Suarez, Fernando and Jantre, Sanket and Kashinath, Karthik and Khairoutdinov, Marat and Kurth, Thorsten and Lutsko, Nicholas and Ma, Po-Lun and Mooers, Griffin and Neelin, J. David and Randall, David and Shamekh, Sara and Taylor, Mark and Urban, Nathan and Yuval, Janni and Zhang, Guang and Pritchard, Mike},
	editor = {Oh, A. and Naumann, T. and Globerson, A. and Saenko, K. and Hardt, M. and Levine, S.},
	year = {2023},
	pages = {22070--22084},
}

@article{lee_representation_2023,
	title = {Representation of atmosphere-induced heterogeneity in land–atmosphere interactions in {E3SM}–{MMFv2}},
	volume = {16},
	testurl = {https://gmd.copernicus.org/articles/16/7275/2023/},
	doi = {10.5194/gmd-16-7275-2023},
	number = {24},
	journal = {Geoscientific Model Development},
	author = {Lee, J. and Hannah, W. M. and Bader, D. C.},
	year = {2023},
	pages = {7275--7287},
}

@article{zangl_icon_2015,
	title = {The {ICON} ({ICOsahedral} {Non}-hydrostatic) modelling framework of {DWD} and {MPI}-{M}: {Description} of the non-hydrostatic dynamical core},
	volume = {141},
	testurl = {https://rmets.onlinelibrary.wiley.com/doi/abs/10.1002/qj.2378},
	doi = {https://doi.org/10.1002/qj.2378},
	number = {687},
	journal = {Quarterly Journal of the Royal Meteorological Society},
	author = {Zängl, Günther and Reinert, Daniel and Rípodas, Pilar and Baldauf, Michael},
	year = {2015},
	testnote = {\_eprint: https://rmets.onlinelibrary.wiley.com/doi/pdf/10.1002/qj.2378},
	pages = {563--579},
}

@article{righi_earth_2020,
	title = {Earth {System} {Model} {Evaluation} {Tool} ({ESMValTool}) v2.0 – technical overview},
	volume = {13},
	testurl = {https://gmd.copernicus.org/articles/13/1179/2020/},
	doi = {10.5194/gmd-13-1179-2020},
	number = {3},
	journal = {Geoscientific Model Development},
	author = {Righi, M. and Andela, B. and Eyring, V. and Lauer, A. and Predoi, V. and Schlund, M. and Vegas-Regidor, J. and Bock, L. and Brötz, B. and de Mora, L. and Diblen, F. and Dreyer, L. and Drost, N. and Earnshaw, P. and Hassler, B. and Koldunov, N. and Little, B. and Loosveldt Tomas, S. and Zimmermann, K.},
	year = {2020},
	pages = {1179--1199},
}

@misc{schroder_combined_2023,
	title = {A combined high resolution global {TCWV} product from microwave and near infrared imagers - {COMBI}},
	testurl = {https://wui.cmsaf.eu/safira/action/viewDoiDetails?acronym=COMBI_V001},
	publisher = {Satellite Application Facility on Climate Monitoring (CM SAF)},
	author = {Schröder, Marc and Danne, Olaf and Falk, Ulrike and Niedorf, Anja and Preusker, Rene and Trent, Tim and Brockmann, Carsten and Fischer, Jürgen and Hegglin, Michaela and Hollmann, Rainer and Pinnock, Simon},
	year = {2023},
	doi = {10.5676/EUM_SAF_CM/COMBI/V001},
}

@article{yao_physics-incorporated_2023,
	title = {A {Physics}-{Incorporated} {Deep} {Learning} {Framework} for {Parameterization} of {Atmospheric} {Radiative} {Transfer}},
	volume = {15},
	testurl = {https://agupubs.onlinelibrary.wiley.com/doi/abs/10.1029/2022MS003445},
	doi = {https://doi.org/10.1029/2022MS003445},
	number = {5},
	journal = {Journal of Advances in Modeling Earth Systems},
	author = {Yao, Yichen and Zhong, Xiaohui and Zheng, Yongjun and Wang, Zhibin},
	year = {2023},
	testnote = {\_eprint: https://agupubs.onlinelibrary.wiley.com/doi/pdf/10.1029/2022MS003445},
	pages = {e2022MS003445},
}

@misc{falcon_pytorch_2019,
	title = {{PyTorch} {Lightning}},
	url = {https://github.com/Lightning-AI/lightning},
	author = {Falcon, William and {The PyTorch Lightning team}},
	month = mar,
	year = {2019},
	doi = {10.5281/zenodo.3828935},
}

@article{hafner_interpretable_2024,
	title = {Interpretable machine learning-based radiation emulation for icon},
	journal = {Authorea Preprints},
	author = {Hafner, Katharina and Iglesias-Suarez, Fernando and Shamekh, Sara and Gentine, Pierre and Giorgetta, Marco A and Pincus, Robert and Eyring, Veronika},
	year = {2024},
	testnote = {Publisher: Authorea},
	}

@article{ukkonen_representing_2024,
	title = {Representing sub-grid processes in weather and climate models via sequence learning},
	journal = {Authorea Preprints},
	author = {Ukkonen, Peter and Chantry, Matthew},
	year = {2024},
	testnote = {Publisher: Authorea},
	}

@article{zhang_global_2023,
	title = {Global {Radiative} {Flux} {Profile} {Data} {Set}: {Revised} and {Extended}},
	volume = {128},
	testurl = {https://agupubs.onlinelibrary.wiley.com/doi/abs/10.1029/2022JD037340},
	doi = {https://doi.org/10.1029/2022JD037340},
	number = {5},
	journal = {Journal of Geophysical Research: Atmospheres},
	author = {Zhang, Yuanchong and Rossow, William B.},
	year = {2023},
	testnote = {\_eprint: https://agupubs.onlinelibrary.wiley.com/doi/pdf/10.1029/2022JD037340},
	pages = {e2022JD037340},
}

@article{wang_stochastic_2016,
	title = {Stochastic convective parameterization improving the simulation of tropical precipitation variability in the {NCAR} {CAM5}},
	volume = {43},
	testurl = {https://agupubs.onlinelibrary.wiley.com/doi/abs/10.1002/2016GL069818},
	doi = {https://doi.org/10.1002/2016GL069818},
	number = {12},
	journal = {Geophysical Research Letters},
	author = {Wang, Yong and Zhang, Guang J. and Craig, George C.},
	year = {2016},
	testnote = {\_eprint: https://agupubs.onlinelibrary.wiley.com/doi/pdf/10.1002/2016GL069818},
	pages = {6612--6619},
}

@article{adam_relation_2016,
	title = {Relation of the double-{ITCZ} bias to the atmospheric energy budget in climate models},
	volume = {43},
	testurl = {https://agupubs.onlinelibrary.wiley.com/doi/abs/10.1002/2016GL069465},
	doi = {https://doi.org/10.1002/2016GL069465},
	number = {14},
	journal = {Geophysical Research Letters},
	author = {Adam, Ori and Schneider, Tapio and Brient, Florent and Bischoff, Tobias},
	year = {2016},
	testnote = {\_eprint: https://agupubs.onlinelibrary.wiley.com/doi/pdf/10.1002/2016GL069465},
	pages = {7670--7677},
}

@article{hwang_link_2013,
	title = {Link between the double-{Intertropical} {Convergence} {Zone} problem and cloud biases over the {Southern} {Ocean}},
	volume = {110},
	testurl = {https://www.pnas.org/doi/abs/10.1073/pnas.1213302110},
	doi = {10.1073/pnas.1213302110},
	number = {13},
	journal = {Proceedings of the National Academy of Sciences},
	author = {Hwang, Yen-Ting and Frierson, Dargan M. W.},
	year = {2013},
	testnote = {\_eprint: https://www.pnas.org/doi/pdf/10.1073/pnas.1213302110},
	pages = {4935--4940},
}

@article{blanchard_mid-level_2021,
	title = {Mid-level convection in a warm conveyor belt accelerates the jet stream},
	volume = {2},
	testurl = {https://wcd.copernicus.org/articles/2/37/2021/},
	doi = {10.5194/wcd-2-37-2021},
	number = {1},
	journal = {Weather and Climate Dynamics},
	author = {Blanchard, N. and Pantillon, F. and Chaboureau, J.-P. and Delanoë, J.},
	year = {2021},
	pages = {37--53},
}

@incollection{lee_future_2021,
	series = {{IPCC} {Assessment} {Reports}},
	title = {Future global climate: scenario-based projections and near-term information},
	booktitle = {Climate {Change} 2021: {The} {Physical} {Science} {Basis}. {Contribution} of {Working} {Group} {I} to the {Sixth} {Assessment} {Report} of the {Intergovernmental} {Panel} on {Climate} {Change}},
	publisher = {Cambridge University Press},
	author = {Lee, J-Y and Marotzke, J and Bala, G and Cao, L and Corti, S and Dunne, JP and Engelbrecht, F and Fischer, E and Fyfe, JC and Jones, C and Maycock, A and Mutemi, J and Niaye, O and Panickal, S and Zhou, T and Christensen, HM},
	year = {2021},
	pages = {553--672},
}

@article{atkinson_ftorch_2025,
	title = {{FTorch}: a library for coupling {PyTorch} models to {Fortran}},
	volume = {10},
	testurl = {https://joss.theoj.org/papers/10.21105/joss.07602},
	doi = {10.21105/joss.07602},
	number = {107},
	journal = {Journal of Open Source Software},
	author = {Atkinson, Jack and Elafrou, Athena and Kasoar, Elliott and Wallwork, Joseph G. and Meltzer, Thomas and Clifford, Simon and Orchard, Dominic and Edsall, Chris},
	month = mar,
	year = {2025},
	pages = {7602},
}

@article{hannah_initial_2020,
	title = {Initial {Results} {From} the {Super}-{Parameterized} {E3SM}},
	volume = {12},
	testurl = {https://agupubs.onlinelibrary.wiley.com/doi/abs/10.1029/2019MS001863},
	doi = {https://doi.org/10.1029/2019MS001863},
	number = {1},
	journal = {Journal of Advances in Modeling Earth Systems},
	author = {Hannah, W. M. and Jones, C. R. and Hillman, B. R. and Norman, M. R. and Bader, D. C. and Taylor, M. A. and Leung, L. R. and Pritchard, M. S. and Branson, M. D. and Lin, G. and Pressel, K. G. and Lee, J. M.},
	year = {2020},
	testnote = {\_eprint: https://agupubs.onlinelibrary.wiley.com/doi/pdf/10.1029/2019MS001863},
	pages = {e2019MS001863},
}

@article{arakawa_multiscale_2011,
	title = {Multiscale modeling of the moist-convective atmosphere — {A} review},
	volume = {102},
	issn = {0169-8095},
	testurl = {https://www.sciencedirect.com/science/article/pii/S0169809511002602},
	doi = {https://doi.org/10.1016/j.atmosres.2011.08.009},
	number = {3},
	journal = {Atmospheric Research},
	author = {Arakawa, A. and Jung, J.-H.},
	year = {2011},
	pages = {263--285},
}

@article{arakawa_cumulus_2004,
	title = {The {Cumulus} {Parameterization} {Problem}: {Past}, {Present}, and {Future}},
	volume = {17},
	testurl = {https://journals.ametsoc.org/view/journals/clim/17/13/1520-0442_2004_017_2493_ratcpp_2.0.co_2.xml},
	doi = {10.1175/1520-0442(2004)017<2493:RATCPP>2.0.CO;2},
	number = {13},
	journal = {Journal of Climate},
	author = {Arakawa, Akio},
	year = {2004},
	testnote = {Place: Boston MA, USA
Publisher: American Meteorological Society},
	pages = {2493 -- 2525},
}

@article{ukkonen2025verticallyrecurrentNN,
    author = {Ukkonen, P. and Chantry, M.},
    title = {Vertically Recurrent Neural Networks for Sub-Grid Parameterization},
    journal = {Journal of Advances in Modeling Earth Systems},
    volume = {17},
    number = {6},
    pages = {e2024MS004833},
    doi = {https://doi.org/10.1029/2024MS004833},
    testurl = {https://agupubs.onlinelibrary.wiley.com/doi/abs/10.1029/2024MS004833},
    eprint = {https://agupubs.onlinelibrary.wiley.com/doi/pdf/10.1029/2024MS004833},
    testnote = {e2024MS004833 2024MS004833},
    year = {2025}
}

@article{klocke2017rediscovery,
  title={Rediscovery of the doldrums in storm-resolving simulations over the tropical Atlantic},
  author={Klocke, Daniel and Brueck, Matthias and Hohenegger, Cathy and Stevens, Bjorn},
  journal={Nature Geoscience},
  volume={10},
  number={12},
  pages={891--896},
  year={2017},
  publisher={Nature Publishing Group UK London}
}

@article {stevens2019AHighAltitudeLongRangeAircraftConfiguredasaCloudObservatoryTheNARVALExpeditions,
      author = "Bjorn Stevens and Felix Ament and Sandrine Bony and Susanne Crewell and Florian Ewald and Silke Gross and Akio Hansen and Lutz Hirsch and Marek Jacob and Tobias Kölling and Heike Konow and Bernhard Mayer and Manfred Wendisch and Martin Wirth and Kevin Wolf and Stephan Bakan and Matthias Bauer-Pfundstein and Matthias Brueck and Julien Delanoë and André Ehrlich and David Farrell and Marvin Forde and Felix Gödde and Hans Grob and Martin Hagen and Evelyn Jäkel and Friedhelm Jansen and Christian Klepp and Marcus Klingebiel and Mario Mech and Gerhard Peters and Markus Rapp and Allison A. Wing and Tobias Zinner",
      title = "A High-Altitude Long-Range Aircraft Configured as a Cloud Observatory: The NARVAL Expeditions",
      journal = "Bulletin of the American Meteorological Society",
      year = "2019",
      publisher = "American Meteorological Society",
      address = "Boston MA, USA",
      volume = "100",
      number = "6",
      doi = "10.1175/BAMS-D-18-0198.1",
      pages=      "1061 - 1077",
      testurl = "https://journals.ametsoc.org/view/journals/bams/100/6/bams-d-18-0198.1.xml"
}

@article {Khairoutdinov2003CloudResolvingModelingoftheARMSummer1997IOPModelFormulationResultsUncertaintiesandSensitivities,
      author = "Marat F. Khairoutdinov and David A. Randall",
      title = "Cloud Resolving Modeling of the ARM Summer 1997 IOP: Model Formulation, Results, Uncertainties, and Sensitivities",
      journal = "Journal of the Atmospheric Sciences",
      year = "2003",
      publisher = "American Meteorological Society",
      address = "Boston MA, USA",
      volume = "60",
      number = "4",
      doi = "10.1175/1520-0469(2003)060<0607:CRMOTA>2.0.CO;2",
      pages=      "607 - 625",
      testurl = "https://journals.ametsoc.org/view/journals/atsc/60/4/1520-0469_2003_060_0607_crmota_2.0.co_2.xml"
}

@Article{egusphere-2025-2473,
    AUTHOR = {M\"uller, W. A. and Lorenz, S. and Pham, T. V. and Schneidereit, A. and Brokopf, R. and Brovkin, V. and Br\"uggemann, N. and Chegini, F. and Dommenget, D. and Fr\"ohlich, K. and Fr\"uh, B. and Gayler, V. and Haak, H. and Hagemann, S. and Hanke, M. and Ilyina, T. and Jungclaus, J. and K\"ohler, M. and Korn, P. and Kornbl\"uh, L. and Kroll, C. and Kr\"uger, J. and Castro-Morales, K. and Niemeier, U. and Pohlmann, H. and Polkova, I. and Potthast, R. and Riddick, T. and Schlund, M. and Stacke, T. and Wirth, R. and Yu, D. and Marotzke, J.},
    TITLE = {The ICON-based Earth System Model for Climate Predictions and Projections (ICON XPP v1.0)},
    JOURNAL = {EGUsphere},
    VOLUME = {2025},
    YEAR = {2025},
    PAGES = {1--60},
    testurl = {https://egusphere.copernicus.org/preprints/2025/egusphere-2025-2473/},
    DOI = {10.5194/egusphere-2025-2473}
}

@article{jumper2021highly,
  title={Highly accurate protein structure prediction with AlphaFold},
  author={Jumper, John and Evans, Richard and Pritzel, Alexander and Green, Tim and Figurnov, Michael and Ronneberger, Olaf and Tunyasuvunakool, Kathryn and Bates, Russ and {\v{Z}}{\'\i}dek, Augustin and Potapenko, Anna and others},
  journal={nature},
  volume={596},
  number={7873},
  pages={583--589},
  year={2021},
  publisher={Nature Publishing Group UK London}
}

@article{bretherton_relationships_2004,
	title = {Relationships between {Water} {Vapor} {Path} and {Precipitation} over the {Tropical} {Oceans}},
	volume = {17},
	testurl = {https://journals.ametsoc.org/view/journals/clim/17/7/1520-0442_2004_017_1517_rbwvpa_2.0.co_2.xml},
	doi = {10.1175/1520-0442(2004)017<1517:RBWVPA>2.0.CO;2},
	number = {7},
	journal = {Journal of Climate},
	author = {Bretherton, Christopher S. and Peters, Matthew E. and Back, Larissa E.},
	year = {2004},
	testnote = {Place: Boston MA, USA
Publisher: American Meteorological Society},
	pages = {1517 -- 1528},
}

@article{holloway_moisture_2009,
	title = {Moisture {Vertical} {Structure}, {Column} {Water} {Vapor}, and {Tropical} {Deep} {Convection}},
	volume = {66},
	testurl = {https://journals.ametsoc.org/view/journals/atsc/66/6/2008jas2806.1.xml},
	doi = {10.1175/2008JAS2806.1},
	number = {6},
	journal = {Journal of the Atmospheric Sciences},
	author = {Holloway, Christopher E. and Neelin, J. David},
	year = {2009},
	testnote = {Place: Boston MA, USA
Publisher: American Meteorological Society},
	pages = {1665 -- 1683},
}

@article{mobis_factors_2012,
	title = {Factors controlling the position of the {Intertropical} {Convergence} {Zone} on an aquaplanet},
	volume = {4},
	issn = {1942-2466},
	doi = {https://doi.org/10.1029/2012MS000199},
	number = {4},
	journal = {Journal of Advances in Modeling Earth Systems},
	author = {Möbis, Benjamin and Stevens, Bjorn},
	year = {2012},
}

@article{JUWELS,
    author = {{J\"{u}lich Supercomputing Centre}},
    title = {{JUWELS Cluster and Booster: Exascale Pathfinder with Modular Supercomputing Architecture at Juelich Supercomputing Centre}},
    journal = {Journal of large-scale research facilities},
    number = {A138},
    volume = {7},
    doi = {10.17815/jlsrf-7-183},
    year = {2021}
}

@article{koldunov2023nextgems,
  title={nextGEMS: output of the model development cycle 3 simulations for ICON and IFS. doi: 10.26050},
  author={Koldunov, Nikolay and K{\"o}lling, Tobias and Pedruzo-Bagazgoitia, Xabier and Rackow, Thomas and Redler, Ren{\'e} and Sidorenko, Dmitry and Wieners, Karl-Hermann and Ziemen, Florian Andreas},
  journal={WDCC/nextGEMS\_cyc3},
  year={2023}
}

@article{ross2023benchmarking,
  title={Benchmarking of machine learning ocean subgrid parameterizations in an idealized model},
  author={Ross, Andrew and Li, Ziwei and Perezhogin, Pavel and Fernandez-Granda, Carlos and Zanna, Laure},
  journal={Journal of Advances in Modeling Earth Systems},
  volume={15},
  number={1},
  year={2023},
  publisher={Wiley Online Library}
}

@article{pritchard2014restricting,
  title={Restricting 32--128 km horizontal scales hardly affects the MJO in the Superparameterized Community Atmosphere Model v. 3.0 but the number of cloud-resolving grid columns constrains vertical mixing},
  author={Pritchard, Michael S and Bretherton, Christopher S and DeMott, Charlotte A},
  journal={Journal of Advances in Modeling Earth Systems},
  volume={6},
  number={3},
  pages={723--739},
  year={2014},
  publisher={Wiley Online Library}
}

@article{sarauer2025physics,
  title={A physics-informed machine learning parameterization for cloud microphysics in ICON},
  author={Sarauer, Ellen and Schwabe, Mierk and Weiss, Philipp and Lauer, Axel and Stier, Philip and Eyring, Veronika},
  journal={Environmental Data Science},
  volume={4},
  pages={e40},
  year={2025},
  publisher={Cambridge University Press}
}

@article{stevens_what_2013,
	title = {What are climate models missing?},
	volume = {340},
	number = {6136},
	journal = {science},
	author = {Stevens, Bjorn and Bony, Sandrine},
	year = {2013},
	note = {Publisher: American Association for the Advancement of Science},
	pages = {1053--1054},
}

@article{satoh_global_2019,
	title = {Global {Cloud}-{Resolving} {Models}},
	volume = {5},
	issn = {2198-6061},
	doi = {10.1007/s40641-019-00131-0},
	number = {3},
	journal = {Current Climate Change Reports},
	author = {Satoh, Masaki and Stevens, Bjorn and Judt, Falko and Khairoutdinov, Marat and Lin, Shian-Jiann and Putman, William M. and Düben, Peter},
	month = sep,
	year = {2019},
	pages = {172--184},
}

@article{krasnopolsky_neural_2008,
	title = {Neural network approach for robust and fast calculation of physical processes in numerical environmental models: {Compound} parameterization with a quality control of larger errors},
	volume = {21},
	number = {2-3},
	journal = {Neural Networks},
	author = {Krasnopolsky, Vladimir M and Fox-Rabinovitz, Michael S and Tolman, Hendrik L and Belochitski, Alexei A},
	year = {2008},
	note = {Publisher: Elsevier},
	pages = {535--543},
}

@article{song2021compound,
  title={Compound parameterization to improve the accuracy of radiation emulator in a numerical weather prediction model},
  author={Song, Hwan-Jin and Roh, Soonyoung and Park, Hyesook},
  journal={Geophysical Research Letters},
  volume={48},
  number={20},
  pages={e2021GL095043},
  year={2021},
  publisher={Wiley Online Library}
}

@software{helgehr_2025_17234569,
  author       = {helgehr},
  title        = {EyringMLClimateGroup/heuer25james\_ml\_convection\_climsim: Beyond the Training Data: Confidence-Guided Mixing of Parameterizations in a Hybrid AI-Climate Model},
  month        = sep,
  year         = 2025,
  publisher    = {Zenodo},
  version      = {v1.0},
  doi          = {10.5281/zenodo.17234569},
  url          = {https://doi.org/10.5281/zenodo.17234569},
  swhid        = {swh:1:dir:076b0042944999b65ee0db850663774d002e6d70;origin=https://doi.org/10.5281/zenodo.17234568;visit=swh:1:snp:4f192472ba0868294391a1d3816adf1f969d10c7;anchor=swh:1:rel:7ba82eb971c8ee1e54f93a8a95326d2a58aedcb5;path=EyringMLClimateGroup-heuer25james\_ml\_convection\_climsim-87a1061},
}

@article{wang2022StableClimate,
  title = {Stable Climate Simulations Using a Realistic General Circulation Model with Neural Network Parameterizations for Atmospheric Moist Physics and Radiation Processes},
  author = {Wang, X. and Han, Y. and Xue, W. and Yang, G. and Zhang, G. J.},
  year = 2022,
  journal = {Geoscientific Model Development},
  volume = {15},
  number = {9},
  pages = {3923--3940},
  issn = {1991-9603},
  doi = {10.5194/gmd-15-3923-2022}
}

@article{han2023EnsembleNeural,
  title = {An {{Ensemble}} of {{Neural Networks}} for {{Moist Physics Processes}}, {{Its Generalizability}} and {{Stable Integration}}},
  author = {Han, Yilun and Zhang, Guang J. and Wang, Yong},
  year = 2023,
  journal = {Journal of Advances in Modeling Earth Systems},
  volume = {15},
  number = {10},
  pages = {e2022MS003508},
  issn = {1942-2466},
  doi = {10.1029/2022MS003508}
}

@article{wood2006RelationshipStratiform,
  title = {On the Relationship between Stratiform Low Cloud Cover and Lower-Tropospheric Stability},
  author = {Wood, Robert and Bretherton, Christopher S.},
  year = 2006,
  journal = {Journal of Climate},
  volume = {19},
  number = {24},
  pages = {6425--6432},
  publisher = {American Meteorological Society},
  address = {Boston MA, USA},
  doi = {10.1175/JCLI3988.1}
}

@article{liu2023UnderstandingPrecipitation,
  title = {Understanding Precipitation Bias Sensitivities in {{E3SM-multi-scale}} Modeling Framework from a Dilution Framework},
  author = {Liu, Nana and Pritchard, Michael S and Jenney, Andrea M and Hannah, Walter M},
  year = 2023,
  journal = {Journal of Advances in Modeling Earth Systems},
  volume = {15},
  number = {4},
  pages = {e2022MS003460},
  publisher = {Wiley Online Library}
}

\end{document}